\documentclass[10pt, pre, amsmath, onecolumn, showpacs, superscriptaddress]{revtex4-1}
\usepackage{graphicx,color}
 \graphicspath{{./}{./figures/}}
\usepackage{ams math}
\usepackage{enumerate}
\usepackage{mathbbol}
\usepackage{amsfonts}
\usepackage{natbib}

\usepackage{bm}

\usepackage{color}
\usepackage[dvipsnames]{xcolor}

\newcommand{\be}{\begin{equation}}
\newcommand{\ee}{\end{equation}}
\newcommand{\bea}{\begin{eqnarray}}
\newcommand{\eea}{\end{eqnarray}}
\newcommand{\beq}{\begin{eqnarray}}
\newcommand{\eeq}{\end{eqnarray}}

\newlength{\bilderlength}

\begin{document}

\title{Non-crossing run-and-tumble particles on a line} 

\author{Pierre Le Doussal}
\affiliation{Laboratoire de Physique de l'Ecole Normale Sup\'erieure, PSL University, CNRS, Sorbonne Universit\'es, 24 rue Lhomond, 75231 Paris, France}
\author{Satya N. \surname{Majumdar}}
\affiliation{LPTMS, CNRS, Univ. Paris-Sud, Universit\'e Paris-Saclay, 91405 Orsay, France}
\author{Gr\'egory \surname{Schehr}}
\affiliation{LPTMS, CNRS, Univ. Paris-Sud, Universit\'e Paris-Saclay, 91405 Orsay, France}

\date{\today}

\begin{abstract} 
We study active particles performing independent run and tumble motion on an infinite 
line with 
velocities $v_0 \sigma(t)$, where $\sigma(t) = \pm 1$ is a dichotomous telegraphic noise
with constant flipping rate $\gamma$. We first consider one particle in the presence
of an absorbing wall at $x=0$ and calculate the probability that it has survived up 
to time $t$ and is at position $x$ at time $t$. We then consider two particles
with independent telegraphic noises and compute exactly the probability that they do not 
cross up to time $t$. Contrarily to the case of passive (Brownian)
particles this two-RTP problem can not be reduced to a single RTP with an 
absorbing wall. Nevertheless, we are able to compute exactly the probability of 
no-crossing
of two independent RTP's up to time $t$ and
find that it decays at large time as $t^{-1/2}$ with an 
amplitude that depends on
the initial condition. The latter allows to define an effective length scale, 
analogous to the so called `` 
Milne extrapolation length'' in neutron scattering, which we demonstrate to be a 
fingerprint of the active dynamics.

\end{abstract}

\maketitle

\section{Introduction}

First-passage properties of a single or multiple Brownian walkers have been studied extensively with
a tremendous range of applications in physics, chemistry, biology, astronomy, and all the way to computer 
science and finance (for reviews see
e.g., Refs.~\cite{Chandra_1943,Redner_book,SM_review,BF_2005,Persistence_review,fp_book_2014} amongst many others).
As a warmup, let us start, for example, with the simple problem of computing the probability that two 
ordinary Brownian particles 
on an infinite line, initially separated by a positive distance, do not cross each other up to time $t$. 
Starting initially at $x(0)$ and $y(0)$, with $x(0)>y(0)$, the positions $x(t)$ and $y(t)$ of the two 
walkers evolve independently by the Langevin dynamics
\begin{equation}
\frac{dx}{dt}= \eta_1(t)\, ; \quad \frac{dy}{dt}= \eta_2(t)\, ,
\label{Browntwo_lange.1}
\end{equation}  
where $\eta_1(t)$ and $\eta_2(t)$ are independent Gaussian white noises with zero mean and correlators
$\langle \eta_i(t)\eta_j(t')\rangle = 2D\, \delta_{i,j}\, \delta(t-t')$ for $i,\, j=1,\,2$.
What is the probability that two particles do not cross each other up to time $t$? 

This classic first-passage question can be solved very easily by considering the relative
coordinate $z(t)= [x(t)-y(t)]/2$ that also evolves as a Brownian motion
\begin{equation}
\frac{dz}{dt}= \eta(t)\,
\label{Brown_rel.1}
\end{equation}
where $\eta(t)= [\eta_1(t)-\eta_2(t)]/2$ is again a Gaussian white noise with zero mean and correlator
{$\langle \eta(t)\eta(t')\rangle= 2 \,D'\, \delta(t-t')$} with an effective diffusion
constant $D'=D/2$. The initial value of $z(t)$ is simply $z_0=[x(0)-y(0)]/2>0$.
Thus, the non-crossing probability of {\em two} particles reduces to the
no zero crossing probability of a {\rm single particle}: what is the probability
that a single Brownian walker, starting initially at $z_0>0$, does not cross the
origin up to time $t$? This resulting single particle problem can be solved quite easily by the
image method~\cite{Chandra_1943,Redner_book,Persistence_review,SM_review}.
Let $P(z,t|z_0)$ denote the probability density that the walker is at 
$z$ at time $t$ starting from $z_0$ at $t=0$ and
that it has not yet crossed the origin during the time interval $[0,t]$. 
Then, $P(z,t|z_0)$ satisfies the diffusion
equation, $\partial_t P= D'\, \partial_z^2 P$, on the semi-infinite line $z\ge 0$ with an absorbing boundary condition
at the wall $z=0$ (origin) and the initial condition $P(z,t=0|z_0)= \delta(z-z_0)$. The exact solution, 
obtained simply via 
the image method, reads
\begin{equation}
P(z,t|z_0)= \frac{1}{\sqrt{4\pi D'\, t}}\, \left[e^{-(z-z_0)^2/{4D' t}}- e^{-(z+z_0)^2/{4D' t}}\right]\, .
\label{Brown_image.1}
\end{equation}
Consequently, the survival probability $S(z_0,t)$, which is obtained by integrating over the final position $z$ at time $t$, is given by
\begin{equation}
S(z_0, t) = \int_0^{\infty} P(z,t|z_0)\, dz = {\rm erf}\left(\frac{z_0}{\sqrt{4D' t}}\right)\, ,
\label{Brown_surv.1}
\end{equation}   
where $D'=D/2$. {In particular}, the survival probability decays algebraically at late times: $S(z_0,t)\sim z_0/\sqrt{\pi\, D'\, t}$ as 
$t\to \infty$.

{ This reduction of the two-body problem to a simpler one-body 
problem with 
an absorbing wall works for 
the ordinary non-interacting Brownian walkers because { the} driving 
noises $\eta_1(t)$ 
and $\eta_2(t)$ are {Gaussian and memoryless}, i.e., delta-correlated. 
Consider again two non-interacting particles moving on a line, but
each of them is driven independently by {\em coloured} noises $\eta_1(t)$
and $\eta_2(t)$ that have a 
finite
memory. When the driving noise has a finite memory, the time evolution
of the position of each walker is non-Markovian. If one is again
interested in the probability of no crossing of the two 
non-interacting non-Markovian walkers,
it is no longer possible to reduce the two-body problem to a one-body 
problem with an absorbing wall as was done for Markovian walkers.
One can still consider a relative coordinate 
$z(t)=
[x(t)-y(t)]/2$, but to study its evolution in time, it is not enough
to consider just the effective driving noise $\eta(t)= 
[\eta_1(t)-\eta_2(t)]/2$. To specify the full temporal evolution
of $z(t)$ one needs to keep track of the individual noises
$\eta_1(t)$ and $\eta_2(t)$.
Consequently, computing the 
non-crossing probability {even for} this simple two-body 
non-interacting but non-Markovian walkers, driven by independent
coloured noises, becomes highly 
nontrivial. The purpose of this paper is to present an 
exact solution of this two-body first-passage 
problem for the so called 
`persistent Brownian motions' that are non-Markovian with a finite 
memory.} 

Our motivation for this work comes from the recent resurgence of interest in persistent Brownian motions in
the context of the dynamics of an active particle, such as the `run-and-tumble particle' (RTP)~\cite{Berg_book,TC_2008}.
Bacterias such as E. Coli move in straight runs, undergo tumbling at the end of a run and choose randomly
a new direction for the next run~\cite{Berg_book,TC_2008}. The tumbling occurs as a Poisson process in time with
rate $\gamma$, i.e, the duration of a run between two successive tumblings is an exponentially distributed
random variable with rate $\gamma$. This dynamics can be modelled by associating an internal orientation
degree of freedom with each particle--the particle moves ballistically in the direction of the current
orientation till the orientation changes. In one dimension, the orientation has only two possibilities, $+$ or $-$.
This RTP dynamics is then an example of persistent Brownian motion (it persists to move in one direction during
a random exponential time and hence retains a finite memory). In one dimension, the position of a single RTP
$x(t)$ then evolves via the Langevin equation
\begin{equation}
\frac{dx}{dt}= v_0\, \sigma(t)\, 
\label{RTP_evol.1}
\end{equation}
where $v_0$ is the intrinsic speed during a run and $\sigma(t)=\pm 1$ is a dichotomous 
telegraphic noise that flips from
one state to another with a constant rate $\gamma$. The effective noise $\xi(t)=v_0\, \sigma(t)$
is coloured which is simply seen by computing its autocorrelation function
\begin{equation}
\langle \xi(t) \xi(t')\rangle= v_0^2\, e^{-2\, \gamma\, |t-t'|}\, .
\label{autocorr.1}
\end{equation}
The time scale $\gamma^{-1}$ is the `persistence' time of a run that encodes the memory
of the noise. In the limit $\gamma\to \infty$, 
$v_0\to \infty$ but keeping the ratio $D= v_0^2/{2\gamma}$ fixed, the noise $\xi(t)$ reduces to
a white noise since
\begin{equation}
\langle \xi(t) \xi(t')\rangle= \frac{v_0^2}{\gamma}\, \left[\gamma\, e^{-2\gamma|t-t'|}\right]
\to 2D\, \delta(t-t')\, .
\label{autocorr.2}
\end{equation}
Thus in this so called `diffusive limit', the persistent random walker $x(t)$ reduces to an
ordinary Brownian motion.

The one dimensional persistent random process or the RTP process in Eq. (\ref{RTP_evol.1})
has been studied extensively in the 
past and many properties are well known including the propagator, the mean exit time
from a confined interval, amongst other observables (see e.g., the reviews~\cite{ML_2017,Weiss_2002}). 
More recent studies include the computation of the mean first-passage time between two fixed points in space 
for a single RTP on a line~\cite{ADP_2014,A2015}, 
and the exact distribution of the first-passage time 
to an 
absorbing wall at the origin~\cite{Malakar_2018} 
in the presence of an additional thermal noise in Eq. (\ref{RTP_evol.1}). 
One dimensional RTP with more than two internal degrees of freedom, 
leading to a generalized telegrapher's equation,
was studied recently in Ref.~\cite{DM_2018}. The first-passage properties of a single RTP was also used as an input
in a recent study of an RTP subject to resetting dynamics~\cite{EM_2018}. Finally, for a single RTP
in a confining harmonic potential in $1$d, while the mean first-passage time was computed long back~\cite{MLW_86},
the full first-passage probability to the origin was computed exactly rather recently~\cite{Dhar_18}. 

Most of these first-passage properties mentioned above concern a single RTP in one dimension. In this paper,
we obtain an exact solution for the non-crossing probability of two independent RTP's on a line. As mentioned earlier,
due to the non-Markovian nature of the driving noise, the two-body first-passage problem can no longer be reduced to a
single RTP in the presence of an absorbing wall (unlike the ordinary or `passive' Brownian case). 
Hence, our result provides an exact first-passage distribution for a
genuine two-body problem and also reveals rather rich and interesting behavior of this two-body 
first-passage probability, as a function of the activity parameter $\gamma$ that characterises the
time-scale of the memory of the driving noise. In the limit $\gamma\to \infty$, $v_0\to \infty$, but with the ratio
$v_0^2/\gamma= 2\, D$ fixed, our results recover the standard Brownian 
result. 
{ Let us remark that recently the two RTP
problem with hardcore interaction on a lattice of finite size $L$ was studied, and the full time-dependent solution
for the probability $P(x,y,t)$ that the two particles are at $x$ and $y$ at time $t$ was computed 
exactly~\cite{SEB_16,SEB_17,Mallmin_18}. However, this 
study differs from our problem in a number of ways. The pair of 
RTP's in Refs.~\cite{SEB_16,SEB_17,Mallmin_18} live on a lattice of finite 
size $L$
and have hard core interaction between them. In contrast, 
the two RTP's in our model live
on the infinite continuous line and are noninteracting. In the lattice 
model the joint probability distribution $P(x,y,t)$ on a finite 
ring of size $L$ reaches a steady state
as $t\to \infty$. In our problem, there is no steady state, and we
are interested in computing the probability of the event that the
two non-interacting RTP's do not cross each other up to time $t$,
which was not addressed in Refs.~\cite{SEB_16,SEB_17,Mallmin_18}.} 

{It is useful to highlight one of the main features of the survival probability that emerges from our study. We first consider a single RTP on a semi-infinite line
in the presence of an absorbing wall at the origin and compute exactly the survival probability $S(x_0,t)$ that the 
particle, starting initially at $x_0>0$,
does not cross the origin up to time $t$. We show that at late times $S(x_0,t)$ decays as
\begin{equation}
S(x_0,t) \simeq \frac{1}{\sqrt{\pi\, D\, t}}\,
\left(x_0+ \xi_{\rm Milne}\right)\, ;  \quad\quad{\rm where}\quad D= \frac{v_0^2}{2\gamma}\, , \quad {\rm and}\quad \xi_{\rm Milne}= 
b_+\, \frac{v_0}{\gamma}
\label{surv_1p.1_intro}
\end{equation}
where $b_+$ is the initial probability that the RTP has a positive velocity $v_0$. This behavior is exactly identical
to that of a passive Brownian motion, with the crucial difference that the amplitude of the $1/\sqrt{\pi D t}$ decay
in the active case approaches a nonzero constant $\xi_{\rm Milne}$ as $x_0\to 0$ (i.e., the initial position
approaches the absorbing wall), while for a passive particle this amplitude vanishes as $x_0\to 0$.
We borrowed the notation $\xi_{\rm Milne}$ from the neutron scattering literature where it appears as the
so called Milne extrapolation length (discussed in detail later).
We find a similar late time behavior for the non-crossing probability $S(z_0,t)$ of two RTP's starting from an initial separtion $2\,z_0$,
\begin{equation}
S(z_0,t) \simeq \frac{1}{\sqrt{\pi\, D'\, t}}\left(z_0+\xi_{\rm Milne}\right)\, ; 
\quad\quad {\rm where}\quad D'=\frac{v_0^2}{4\gamma}\, , \quad {\rm and} \quad \xi_{\rm Milne}= \frac{v_0}{2\gamma}\,(1+ b_{+-}-b_{-+})
\label{2_surv_rep.2_intro}
\end{equation}
where $b_{\sigma_1,\sigma_2}$ denote the initial probability that the first particle starts with a velocity $\sigma_1\, v_0$ while the
second particle with velocity $\sigma_2\, v_0$. In this case also, the amplitude of the $1/\sqrt{\pi D' t}$ late time decay approaches a
nonzero constant $\xi_{\rm Milne}$ as in Eq. (\ref{2_surv_rep.2_intro}) when $z_0\to 0$, 
in contrast to the case of two passive Brownian particles where
this amplitude vanishes when $z_0\to 0$. Thus the amplitude of the late time decay of the survival probability 
carries an important fingerprint of the activeness of the particles: while for active particles the Milne extrapolation length
is nonzero $\xi_{\rm Milne}>0$, for passive particles $\xi_{\rm Milne}=0$ identically.}

The rest of our paper is organised as follows. In Section II, we 
consider a single RTP on the semi-infinite line with an absorbing wall at the origin and compute
exactly the probability $P(x,t|x_0)$ that the walker reaches the position $x$ at time $t$, starting from $x_0$,
and does not cross the origin up to $t$. By integrating over the final position $x$, we recover some of the
known results for the survival probability of a single RTP. However, our results for the spatial 
probability density $P(x,t|x_0)$
contain more information than just the survival probability. 
We show that our method 
can be generalised to the two-particle case and allows us to obtain the exact solution for the 
two-particle case--this is presented in Section III. Finally, we present a summary, conclusion and open problems in
Section IV. Some details on the exact inversion of a number of Laplace transforms are provided in
three Appendices. 

\section{A single RTP in the presence of an absorbing wall at the origin}

We start with a single RTP on a line, whose position $x(t)$ at time $t$ evolves stochastically via
Eq. (\ref{RTP_evol.1}) where $\sigma(t)=\pm 1$ is the telegraphic noise. The noise $\sigma(t)$ changes 
from its current state (say $`+1'$) to the opposite state $`-1'$ (and vice versa) at a constant rate $\gamma$, 
independently of the particle's position. In addition, there is an absorbing wall at the origin $0$. If the
particle crosses the origin, it dies. The RTP starts initially at $x_0>0$ and with its initial 
{internal} state $\sigma(0)=+1$ with
probability $b_+$ and $\sigma(0)=-1$ with probability $b_-$, with $b_++b_-=1$ (we will
focus mostly on the case $b_+=b_-=\frac{1}{2}$). Let $P_{\pm}(x,t)$ denote the probability density
that the particle survives up to time $t$ {\em and} arrives at the position $x$ at time $t$ with its 
{internal} state 
$\sigma(t)=\pm 1$ respectively. For simplicity of notations, we suppress the $x_0$ dependence
of $P(x,t)$ for the moment and will re-instate explicitly the $x_0$ dependence whenever needed. 
Let us also define the total probability density as
\begin{equation}
P(x,t) = P_+(x,t)+ P_{-}(x,t)\, .
\label{Ptotal.0}
\end{equation}

It is easy to derive the Fokker-Planck equations governing the time evolution
of $P_{\pm} (x,t)$ in $x\ge 0$. Consider {the} time evolution from $t$ to $t+dt$. Then
\bea
&& P_+(x, t+dt)  =  [1-\gamma\, dt]\, P_+(x- v_0\, dt, t) + \gamma\, dt\, P_{-}(x,t) \label{P+evol.0} \\
&& P_{-}(x,t+dt)  =  [1-\gamma\, dt]\, P_-(x+ v_0\, dt, t) + \gamma\, dt\, P_{+}(x,t)\, . \label{P-evol.0}
\eea
This is easy to understand. With probability $(1-\gamma\, dt)$ the noise does not change sign during $dt$--hence
if the particle is to arrive at $x$ at $t+dt$ without changing noise from $+1$, it must have been at
$x-v_0 dt$ at time $t$ with internal state $+1$. This explains the first term on the right hand side 
(rhs) of Eq. (\ref{P+evol.0}).
On the other hand, the {internal state} flips with probability $\gamma\, dt$ in time $dt$ 
during which the particle position
does not change. Hence, the particle can be at $x$ at $t+dt$ with internal state $+1$ 
if it was at $x$ at time $t$ with internal state $-1$ ---this event happens with 
probability $\gamma\, dt$, explaining the second term on the rhs of Eq. (\ref{P+evol.0}). 
Similar reasonings lead to the  
second equation (\ref{P-evol.0}) for $P_{-}(x,t)$. Taking $dt\to 0$ limit leads to the pair of Fokker-Planck
equations
\bea
&& \partial_t P_+ = - v_0 \partial_x P_+ -  \gamma P_+ + \gamma P_-  \label{P+evol.1}\\
&& \partial_t P_- = v_0 \partial_x P_- -  \gamma P_- + \gamma P_+ \, . 
\label{P-evol.1}
\eea
The first terms in both equations describe the advection terms caused by the ballistic motion of the RTP during a `run',
while the last two terms (in each equation)  describe the loss and gain incurred due to the {change} of sign by the
driving telegraphic noise. These equations evolve on the semi-infinite line $x\ge 0$ starting from the initial
condition
\begin{equation}
P_{+}(x,0)= b_+ \, \delta(x-x_0)\quad {\rm and}\quad P_{-}(x,0)= b_- \, \delta(x-x_0)\, .
\label{init_cond.1}
\end{equation}

Finally, we need to specify the boundary condition at $x=0$ and $x\to \infty$. As $x\to \infty$, clearly
${P_{\pm }}(x\to \infty, t)=0$ since the RTP, irrespective of its internal state, can not reach $\infty$ in a finite time $t$,
starting from a finite $x_0>0$. In contrast, the absorbing boundary condition at $x=0$ is more tricky to write down.
This boundary condition can be deduced by considering the microscopic time evolution of a trajectory starting
at $x=0$. Consider first Eq. (\ref{P+evol.0}) and set $x=0$
\begin{equation}
P_+(0, t+dt)  =  [1-\gamma\, dt]\, P_+(-v_0\, dt, t) + \gamma\, dt\, P_{-}(0,t)\, .
\label{P+evol.2} 
\end{equation}
Since, by definition, the particle dies when it crosses the origin, there is no particle 
at $x= -v_0\, dt< 0$ at time $t$. 
Consequently, the first term on the rhs of Eq. (\ref{P+evol.2}) is identically $0$. Now, taking $dt\to 0$ limit, we see
that the appropriate boundary condition at $x=0$ is
\begin{equation}
P_+(x=0,t)=0\, .
\label{P+_bc.1}
\end{equation}
We can repeat the same exercise for $P_{-}(x=0,t)$. Putting $x=0$, taking the $dt\to 0$ limit and using $P_+(0,t)=0$, we
arrive at
\begin{equation}
{\partial_t} P_{-}(0,t)= v_0\, \partial_x P_{-}\big|_{x=0} - \gamma\, P_{-}(0,t)\, .
\label{P-_bc.1}
\end{equation}
In other words, it just gives back the Fokker-Planck equation (\ref{P-evol.1}) at $x=0$, and does not provide 
any extra boundary condition. Hence, we see that $P_{+}(0,t)=0$, while $P_{-}(0,t)$ is 
unspecified and its value at $x=0$ is decided by the solution itself (there is no additional information). This `single' boundary
condition is a typical hallmark of persistent Brownian motion. We will see later {that}, just this single boundary 
condition at $x=0$ for $P_{+}(x,t)$, in addition to those at $x\to \infty$, is sufficient to determine uniquely
both $P_{\pm}(x,t)$ at all times $t$.  

To solve the pair of Fokker-Planck equations (\ref{P+evol.1}) and (\ref{P-evol.1}), it is convenient first to
define their Laplace transforms in space
\begin{equation}
{\tilde P}_{\pm}(p,t)= \int_0^{\infty} P_{\pm}(x,t)\, e^{-p\, x}\, dx \,,
\label{px_laplace.1}
\end{equation}
with the initial conditions, using Eq. (\ref{init_cond.1})
\begin{equation}
{\tilde P}_{\pm}(p, t=0)= b_{\pm}\, e^{-p x_0}\, .
\label{px_laplace_init.1}
\end{equation}
Taking Laplace transforms of Eqs. (\ref{P+evol.1}) and (\ref{P-evol.1}) 
with respect to $x$ gives
\bea
&& \partial_t {\tilde P}_{+}(p,t)= -(\gamma+v_0\,p)\, {\tilde P}_+ + \gamma\, {\tilde P}_{-}+ v_0\, P_{+}(x=0,t)\, \label
{P+evol.4} \\
&& \partial_t {\tilde P}_{-}(p,t)= -(\gamma-v_0\,p)\, {\tilde P}_{-}+ \gamma\, {\tilde P}_{+}- v_0\, P_{-}(x=0,t)\,.
\label{P-evol.4}
\eea
We then take the Laplace transforms with respect to $t$
\begin{equation}
{\cal P}_{\pm}(p,s)= \int_0^{\infty} {\tilde P}_{\pm}(p,t)\, e^{-s\,t}\, dt= \int_0^{\infty}dt\, e^{-s\,t}\, 
\int_0^{\infty} dx\,
e^{-p\, x}\, P_{\pm}(x,t)\, ,
\label{double_laplace.1}
\end{equation}
which gives, from Eqs. (\ref{P+evol.4}) and (\ref{P-evol.4}) and using the initial conditions (\ref{px_laplace_init.1}),
\bea
&& (s+\gamma+ v_0\, p) \, {\cal P}_+(p,s)- \gamma\, {\cal P}_{-}(p,s)= b_+ \, e^{-p\, x_0}+v_0\, q_+(0,s) 
\label{P+evol.5}\\
&& (s+\gamma- v_0\, p) \, {\cal P}_-(p,s)- \gamma\, {\cal P}_{+}(p,s) = b_- \, e^{-p\, x_0}-v_0\, q_-(0,s)\, \,
\label{P-evol.5}
\eea
where we have defined the boundary condition dependent terms
\begin{equation}
q_{\pm}(0,s)= \int_0^{\infty} P_{\pm}(0, t)\, e^{-s\, t}\, dt\, .
\label{boundary_unknowns.0}
\end{equation}
Note that, from the boundary condition (\ref{P+_bc.1}), we have $q_{+}(0,s)=0$ identically. Only
$q_{-}(0,s)$ remains unknown and yet to be fixed.

The pair of linear equations (\ref{P+evol.5}) and (\ref{P-evol.5}) can be easily 
solved by inverting the $(2\times 2)$ matrix   
\be
\begin{pmatrix}
{\cal P}_+ \\
{\cal P}_{-}
\end{pmatrix}
=  
\begin{pmatrix}
s+\gamma+ v_0\, p  & - \gamma \\
-\gamma & s+\gamma-v_0\, p  
\end{pmatrix}^{-1} \left(
\begin{pmatrix}
0 \\
- v_0 q_{-}(0,s)
\end{pmatrix}
+ 
e^{- p\, x_0} 
\begin{pmatrix}
b_+\\
b_-
\end{pmatrix}
 \right) \;. \label{eqmatrix} 
\ee 

{While further computations can be carried out straightforwardly
for general inhomogeneous initial condition, i.e, for arbitrary $b_{+}$ and
$b_{-}=1-b_{+}$, 
it turns out that the intermediate
steps leading to the final result are somewhat simpler to display
for the homogeneous case $b_{\pm}=1/2$. Hence, below we first detail
the intermediate steps for the homogeneous case and later we only 
display the final results for the generic inhomogeneous case. The
intermediate steps are similar in both cases.}

\vskip 0.4cm

\noindent {\it Homogeneous initial condition $b_{\pm}=1/2$}.
Setting $b_{\pm}=1/2$ in Eq. (\ref{eqmatrix}),
inverting the $(2\times 2)$ matrix explicitly and adding {the two 
equations for ${\cal 
P}_+(p,s)$ and ${\cal P}_-(p,s)$}, we get
\bea
{\cal P}(p,s)=
{\cal P}_+(p,s)+{\cal P}_-(p,s)= \frac{v_0\, q_-(0,s)\, \left(2 \gamma + s+ p\, v_0\right)-(s+2 \gamma)\, 
e^{-p\, x_0}}{v_0^2\, p^2-s^2-2\, \gamma\,  s} \;,
\label{Total_calP.1}
\eea 
where $q_{-}(0,s)$ is yet to be determined. To fix $q_{-}(0,s)$, we first locate the poles of the
rhs of Eq. (\ref{Total_calP.1}) in the complex $p$ plane
\be
v_0^2\, p^2-s^2-2\, \gamma\,  s = 0 \quad {\Longrightarrow} \quad p^*_{\pm} = \pm \frac{\sqrt{s} \sqrt{2 \gamma +s}}{v_0}\, .
\label{poles.1}
\ee 
Note that $p^*_+>0$. Clearly, if the residue at this pole $p^*_+$ 
is nonzero, this would mean
that upon inversion with respect to $p$, the Laplace transform with respect to time, $\int_{0}^{\infty}
P(x,t)\, e^{-s\, t}\, dt$,
would diverge as $\sim e^{p^*\, x}$ as $x\to \infty$. This is however forbidden by the boundary condition
that $P(x\to \infty,t)=0$. Hence the numerator of the rhs of Eq. (\ref{Total_calP.1}) must vanish
at $p=p^*_+$ (so that there is no pole at $p^*_+$), leading to a unique 
value of $q_{-}(0,s)$ 
\begin{equation}
q_{-}(0,s)= \int_0^{\infty} P_{-}(0,t)\, e^{-s\,t}\, dt  = 
\frac{\sqrt{s+2\,\gamma}}{v_0\,\left(\sqrt{s}+\sqrt{s+2\,\gamma}\right)}\, 
e^{-\sqrt{ \frac{s(s+2\,\gamma)}{v_0^2}}\, x_0}\, .
\label{p-0s.1}
\end{equation}
This pole-cancelling mechanism to fix an unknown boundary term has been used before in
other contexts such as in the exact solution of a class of mass transport models~\cite{RM_00,RM_00.1}. 
The result in Eq. (\ref{p-0s.1}) clearly shows that while $P_+(0,t)=0$ for all $t$, $P_{-}(0,t)$ is nonzero
and is determined by the dynamics itself. Since, $P_+(0,t)=0$, the total probability density at the
wall (starting from $x_0$) is $P(0,t|x_0)= P_{-}(0,t)$ with Laplace transform
\begin{equation} 
\int_0^{\infty} P(0,t|x_0)\, e^{-s\,t}\, dt  =
\frac{\sqrt{s+2\,\gamma}}{v_0\,\left(\sqrt{s}+\sqrt{s+2\,\gamma}\right)}\,
e^{-\sqrt{ \frac{s(s+2\,\gamma)}{v_0^2}}\, x_0}\, .
\label{ptotal0t.1}
\end{equation}
Amazingly, this Laplace transform can be exactly inverted (see Appendix A) giving 
\begin{eqnarray}
&& P(0,t|x_0) =  \frac{\gamma\, e^{-\gamma t}}{2v_0}
\left[\frac{x_0}{x_0+v_0\,t}\, I_0(\rho)
+ \frac{1}{\rho}
\left(\frac{v_0\,t-x_0}{v_0\,t+x_0}+ \frac{\gamma\, x_0}{v_0}\right)\,
I_1(\rho) \right]\, \theta(v_0\,t-x_0)
+ \frac{e^{-\gamma\,t}}{2}\, \delta(v_0\,t-x_0) \nonumber \\
&& {\rm with}\quad  \rho= \frac{\gamma}{v_0}\sqrt{v_0^2\,t^2-x_0^2}\, .
\label{exact_inv.1}
\end{eqnarray}
Here $I_0(z)$ and $I_1(z)$ are modified Bessel functions. The last term corresponds to 
particles of velocities $-v_0$ which have not changed their state since $t=0$.
The asymptotic behaviors for small and large 
$t$, with fixed $x_0$, are given by  
\begin{eqnarray}
P(0,t|x_0) \approx \begin{cases}
&\frac{1}{2} \, \delta(x_0)\;, \quad\quad\quad\,\hspace*{2.5cm} {\rm as}\quad t\to 0  \\
\\
&\dfrac{1}{\sqrt{2\pi\, \gamma\, v_0^2}}\,\left( \frac{1}{2} + 
\dfrac{\gamma\,x_0}{v_0}\right)\,
\dfrac{1}{t^{3/2}}\;, \quad {\rm as}\quad t\to \infty\, .
\end{cases}
\label{p-0t_asymp.1}
\end{eqnarray}
Thus interestingly, $P(0,t|x_0)$ has a slow algebraic decay $\sim t^{-3/2}$ at late times. 
It can also be seen from the term $\sim \sqrt{s}$ in the small $s$ expansion of \eqref{ptotal0t.1}.

It is also instructive to investigate $P(0,t|x_0)$ in Eq. (\ref{exact_inv.1}) for fixed time $t$, but
in the diffusive
limit $v_0\to \infty$, $\gamma\to \infty$ while keeping $v_0^2/\gamma=2D$ fixed. In this limit,
\begin{equation}
\rho= \frac{\gamma}{v_0}\sqrt{v_0^2\,t^2-x_0^2} \to \gamma\, t - \frac{x_0^2}{4Dt} +\dots
\label{diff_lim.1}
\end{equation}
Consequently, Eq. (\ref{exact_inv.1}) reduces to
\begin{equation}
P(0,t|x_0) \approx \frac{1}{v_0}\, \frac{x_0}{\sqrt{4\,\pi\, D\,t^3}}\, e^{-x_0^2/{4Dt}}\, .
\label{diff_lim.2}
\end{equation}
Thus, the probability density at the origin vanishes as $1/v_0$ as $v_0\to \infty$. This
is expected since in the diffusive limit, the probability density at the {absorbing} origin vanishes
identically. For an RTP, this density at the origin is nonzero at finite time $t$ due to
the finite nonzero density of the left movers (i.e., $P_{-}(0,t)$ ). An alternative way to
arrive at the same limiting form in Eq. (\ref{diff_lim.2}) is as follows. We keep $v_0$ and $\gamma$ fixed, 
but take $x_0\to \infty$, $t\to \infty$
with $x_0/\sqrt{t}$ fixed. Analysing Eq. (\ref{exact_inv.1}) in this scaling limit, one arrives at the same 
result (\ref{diff_lim.2}) with $D=v_0^2/{2\gamma}$. 

One notes that Eq. \eqref{diff_lim.2} in the limit of large $t$ gives precisely the second term in the 
large $t$ decay in the second line of \eqref{p-0t_asymp.1} using $D=v_0^2/{2\gamma}$. 
The first term in \eqref{p-0t_asymp.1} is however specific to the active system: we 
observe that the
factor $1/2$ in the first term is precisely 
the probability that the particle has velocity $+ v_0$ at $t=0$ (for this homogeneous
initial condition). 

\begin{figure}
\includegraphics[width = 0.65 \linewidth]{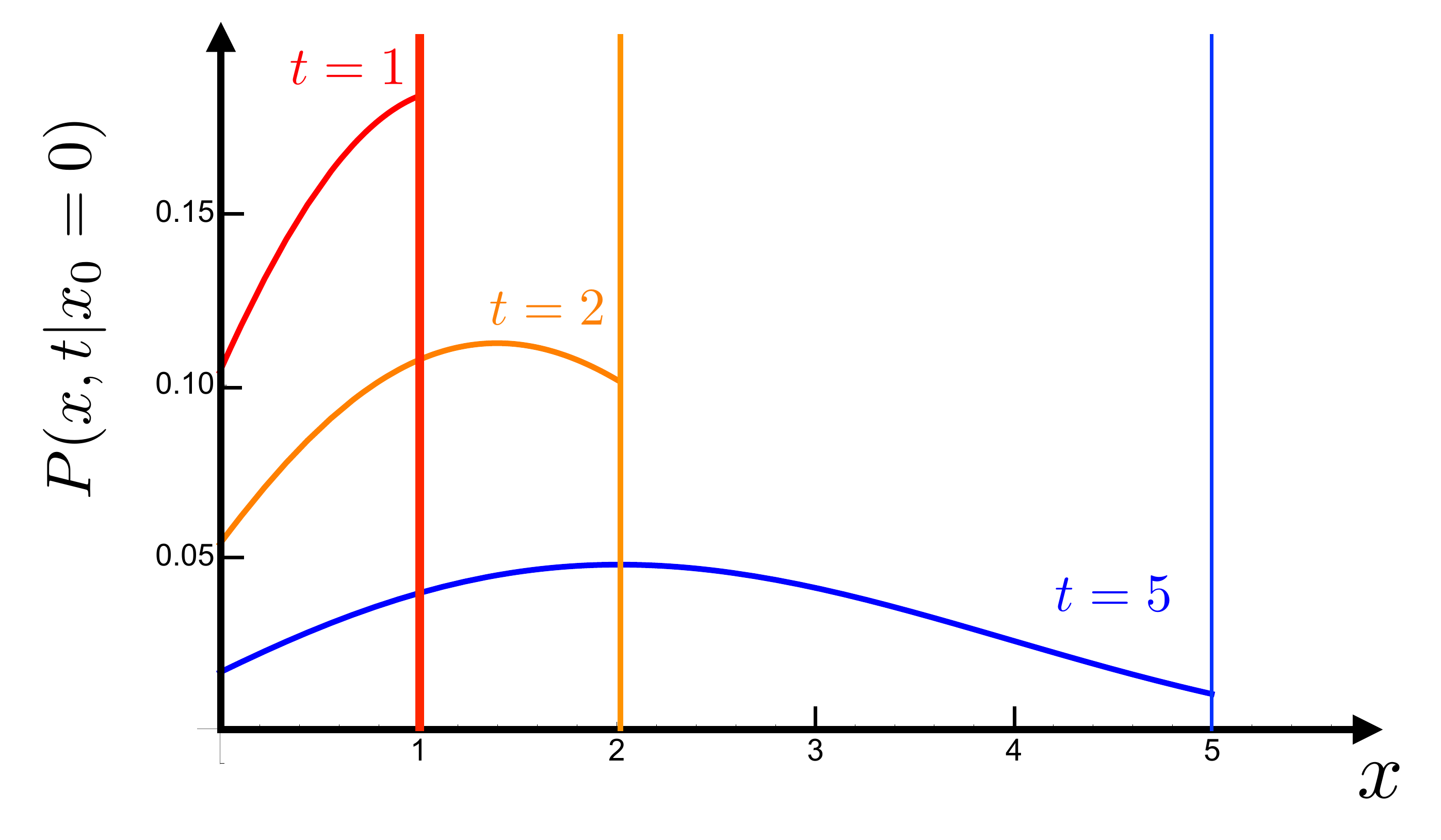}
\caption{The density $P(x,t|x_0=0)$  in Eq. (\ref{exact_inv.2}) is plotted as a function of $x$ for {three different
times $t=1$ (red), $t=2$ (orange) and $t=5$ (blue) with parameter values $v_0=1$ and $\gamma=1$ and symmetric initial conditions $b_\pm = 1/2$. For these parameter values, the range of $x$ is over
$x\in [0,t]$ and at $x=t$, there is a delta function (indicated by the colored vertical lines) with amplitude $e^{-t}/2$ which corresponds to a right moving particle which has not tumbled up to time $t$. This delta peak at $x=t$ damps down exponentially fast with time $t$ (which is sketched by a thinner vertical line as time increases).}
}
\label{fig_Pxt5}
\end{figure}

Inserting $q_{-}(0,s)$ from Eq. (\ref{p-0s.1}) into Eq. (\ref{Total_calP.1}) we get the 
double Laplace transform
of the total probability density $P(x,t)$ 
\bea
{\cal P}(p,s)={\cal P}_+(p,s)+{\cal P}_-(p,s)= 
\frac{ \sqrt{s+2\,\gamma}}{s(s+2\gamma)-v_0^2\, p^2}\, \left[
\sqrt{s+2\gamma}\, e^{-p\, x_0}-
\frac{v_0\,p+s+2\gamma}{\sqrt{s+2\,\gamma}+\sqrt{s}}\, e^{-\sqrt{ \frac{s(s+2\,\gamma)}{v_0^2}}\, x_0}
\right]\, .
\label{Total_calP.2}
\eea
From this exact double Laplace transform, one can easily compute the survival probability $S(x_0,t)$ of the 
RTP up to time $t$, starting from $x_0$. This is obtained by integrating over the final position:
$S(x_0,t)= \int_0^{\infty} P(x,t)\, dx$. Consequently, one gets 
\begin{equation}
\int_0^{\infty} S(x_0,t)\, e^{-s\,t}\,dt= {\cal P}(p=0,s)= \frac{1}{s}\left[1- 
\frac{\sqrt{s+2\gamma}}{\sqrt{s+2\,
\gamma}+ \sqrt{s}}\, e^{-\sqrt{ \frac{s(s+2\,\gamma)}{v_0^2}}\, x_0}\right]\,  \;.
\label{surv.1}
\end{equation}
Interestingly, by comparing this result (\ref{surv.1}) with the result obtained before 
for $P(0,t|x_0)$ in Eq. (\ref{ptotal0t.1}), we find that the first-passage probability 
to the origin $\partial_t S(x_0,t)$ is given by
\begin{equation}
\partial_t S(x_0,t) = - v_0 \, P(0,t|x_0) 
\label{relation.1} \;.
\end{equation}
This can be understood as follows. Defining a probability current $J(x,t)$ such that 
$\partial_t P(x,t)=- \partial_x J(x,t)$, we see from 
Eqs. (\ref{P+evol.1}) and  (\ref{P-evol.1}) that 
$J(x,t) = v_0 [P_+(x,t) - P_-(x,t)]$. In particular, the current at $x=0$ is 
$J(x=0,t) = - v_0 P_-(0,t)$ since $P_+(0,t) = 0$ [see Eq. (\ref{P+_bc.1})]. Integrating over space, one thus has $\partial_t S(x_0,t) = - [J(x,t)]_0^{+\infty} = J(0,t) = - v_0 P_-(0,t)$, which, 
by further using that $P(0,t|x_0) = P_+(0,t|x_0) + 
P_-(0,t|x_0) = P_-(0,t|x_0)$, yields the relation in Eq. (\ref{relation.1}).
{Using the asymptotic decay of $P(0,t|x_0)$ for large $t$ from Eq. (\ref{p-0t_asymp.1}) on the right hand side
of Eq. (\ref{relation.1}) and integrating over $t$, we get the large $t$ decay of the survival probability $S(x_0,t)$ for fixed $x_0$ 
\begin{equation}
S(x_0,t) \simeq \frac{1}{\sqrt{\pi\, D\, t}}\, 
\left(x_0+ \frac{v_0}{2\gamma}\right)\, ;  \quad\quad{\rm where}\quad D= \frac{v_0^2}{2\gamma}\, .
\label{surv_1p.1}
\end{equation}
A similar result holds for more general inhomogeneous initial condition as we show later.}

The result in Eq. (\ref{surv.1}) for the homogeneous initial condition
coincides with the known result on survival probability that was originally deduced by using a backward 
Fokker-Planck approach~\cite{Malakar_2018,EM_2018}. Here we used a forward Fokker-Planck method that gave us
access to a more general quantity, namely the joint probability $P(x,t)$ that the particle survives up to $t$
and arrives at $x$ at time $t$. To the best of our knowledge, we have not come across, in the literature,
the explicit double Laplace transform
of the joint probability in Eq. (\ref{Total_calP.2}). This result (\ref{Total_calP.2}) simplifies a bit
for the special initial position $x_0=0$
\begin{equation}
{\cal P}(p,s|x_0=0)= \frac{ \sqrt{s+2\gamma}}{\left(\sqrt{s+2\gamma}+\sqrt{s}\right)\, \left(v_0\,p +
\sqrt{s(s+2\gamma)}\right)}\, .
\label{Total_calP_x0.0}
\end{equation} 
Inverting trivially with respect to $p$ we get
\begin{equation}
\int_0^{\infty} P(x,t|x_0=0)\, e^{-s\,t}\,dt=   \frac{\sqrt{s+2\gamma}}{v_0\, \left(\sqrt{s+2\gamma}+\sqrt{s}\right)}\,
e^{-\sqrt{ \frac{s(s+2\,\gamma)}{v_0^2}}\, x}\, .
\label{Pxt_lt.1}
\end{equation}
Comparing the rhs of Eqs. (\ref{Pxt_lt.1}) and (\ref{p-0s.1}), we notice the identity 
valid at all times 
\begin{equation}
P(x,\, t|x_0=0)= P(0,\, t|x_0=x)\, ,
\label{time_reversal.1}
\end{equation}
which expresses the time-reversal symmetry valid in this special case of
homogeneous initial condition $b_{\pm}=1/2$. Thus, for this initial 
condition $x_0=0$, 
we can explicitly
invert the Laplace transform (as in Eq. (\ref{exact_inv.1})) to obtain the total
probability density $P(x,t|x_0=0)$ 
\begin{eqnarray}
&& P(x,t|x_0=0) =  \frac{\gamma\, e^{-\gamma t}}{2v_0}\left[\frac{x}{x+v_0\,t}\, I_0(\rho)
+ \frac{1}{\rho}
\left(\frac{v_0\,t-x}{v_0\,t+x}+ \frac{\gamma\, x}{v_0}\right)\, I_1(\rho) \right]\, \theta(v_0\,t-x)
+ \frac{e^{-\gamma\,t}}{2}\, \delta(v_0\,t-x) \nonumber \\
&& {\rm with}\quad  \rho= \frac{\gamma}{v_0}\sqrt{v_0^2\,t^2-x^2}\, .
\label{exact_inv.2}
\end{eqnarray}
A plot of $P(x,t|x_0=0)$ is provided { in} Fig. \ref{fig_Pxt5}. 
The result in Eq. (\ref{exact_inv.2})
can be cast in a scaling form in terms of two dimensionless scaling 
variables: $z=x/{(v_0\, t)}$ and 
$T=\gamma\, t$. One gets
\begin{equation}
P(x,t|x_0=0)= \frac{\gamma}{2v_0}\, F\left(\frac{x}{v_0\, t},\, \gamma\, t\right)
\label{Px_scaling.1}
\end{equation}
where the scaling function $F(z,T)$ is given by
\begin{equation}
F(z,T)= e^{-T}\left[\frac{z}{z+1}\,I_0\left(T\,\sqrt{1-z^2}\right)+ 
\frac{1}{T\sqrt{1-z^2}}\left(\frac{1-z}{1+z}+z\, T\right)\, 
I_1\left(T\, \sqrt{1-z^2}\right)\right]\theta(1-z)+ \frac{1}{2}\, e^{-T}\, \delta(1-z)\, .
\label{Px_scaling.2}
\end{equation}

Finally, we remark that in the diffusive limit $v_0\to \infty$, $\gamma\to \infty$ while keeping the ratio
$v_0^2/\gamma=2\, D$ fixed, Eq. (\ref{Total_calP.2}) reduces to
\be
{\cal P}(p,s) \simeq \frac{e^{-\frac{\sqrt{s} x_0}{\sqrt{D}}}-e^{-p\, x_0}}{D p^2-s}\, .
\label{diff_limit.1}
\ee 
This double transform can be easily inverted to give
\begin{equation}
P(x,t|x_0)= \frac{1}{\sqrt{4\pi\, D\, t}}\, \left[e^{-(x-x_0)^2/{(4\, D\, t)}}- 
e^{-(x+x_0)^2/{(4\, D\, t)}}\right]\, .
\label{image_reduction.1}
\end{equation}
{This is precisely the image solution of an ordinary Brownian motion with an absorbing wall
at the origin~\cite{Chandra_1943,Redner_book,Persistence_review}.
Hence, we verify that in this diffusive limit, the RTP behaves as an ordinary `passive' Brownian motion 
with diffusion constant $D$, as expected.}

\vskip 0.4cm

\noindent {\it Inhomogeneous initial condition.} The technique used above for
the homogeneous case $b_{\pm}=1/2$ generalises, in a straightforward manner, 
to the generic inhomogeneous initial condition with arbitrary $b_+$ and 
$b_{-}=1-b_+$. Without repeating the intermediate steps, we just provide
the main results here. The analogue of Eq. (\ref{exact_inv.1}) for $P(0,t|x_0)$, for 
arbitrary $b_+$, reads
\begin{eqnarray}
&& P(0,t|x_0) =  \frac{\gamma\, e^{-\gamma t}}{v_0}
\left[\frac{b_+ x_0}{x_0+v_0\,t}\, I_0(\rho)
+ \frac{1}{\rho}
\left(b_+ \frac{v_0\,t-x_0}{v_0\,t+x_0}+ b_- 
\frac{\gamma\, x_0}{v_0}\right)\, I_1(\rho) \right]\, \theta(v_0\,t-x_0) 
+ b_- e^{-\gamma\,t}\, \delta(v_0\,t-x_0) \nonumber \\
&& {\rm with}\quad  \rho= \frac{\gamma}{v_0}\sqrt{v_0^2\,t^2-x_0^2}\, . 
\label{exact_inv.1.inh}
\end{eqnarray}
Consequently, its asymptotic behaviors for small and large $t$, for fixed $x_0$,
are given by
\begin{eqnarray}
P(0,t|x_0) \approx \begin{cases}
&b_- \, \delta(x_0)\;, \quad\quad\quad\,\hspace*{2.5cm} 
{\rm as}\quad t\to 0  \\
\\
&\dfrac{1}{\sqrt{2\pi\, \gamma\, v_0^2}}\,
\left( b_+ + \dfrac{\gamma\,x_0}{v_0}\right)\,
\dfrac{1}{t^{3/2}}\;, \quad {\rm as}\quad t\to \infty\, .
\end{cases}
\label{p-0t_asymp.1.inh}
\end{eqnarray}
Note again that the first term in the large $t$ asymptotics (in the second line
of Eq. (\ref{p-0t_asymp.1.inh})) is proportional to $b_+$, i.e., the probability
that initially the RTP has a velocity $+v_0$.

The survival probability $S(x_0,t)$, for general $b_{+}$, turns out to be exactly the same
as in Eq. (\ref{surv.1}) for the homogeneous case, up to an overall factor $2\,b_+$ and we get
\begin{equation}
\int_0^{\infty} S(x_0,t)\, e^{-s\,t}\,dt= 
{\cal P}(p=0,s)= \frac{2\, b_+}{s}\left[1-
\frac{\sqrt{s+2\gamma}}{\sqrt{s+2\,
\gamma}+ \sqrt{s}}\, e^{-\sqrt{ \frac{s(s+2\,\gamma)}{v_0^2}}\, 
x_0}\right]\,  \;.
\label{surv.1_inh}
\end{equation}
For instance in the special case $x_0=0$ the Laplace inversion gives
\be
S(0,t) = b_+ e^{-\gamma t} ( I_0(\gamma t) + I_1(\gamma t)) 
\label{surv_0t.1}
\ee 
for $t \geq 0^+$, 
noting that $S(0,0)=1$ (by definition), but $S(0,0^+)= 1- b_-=b_+$ from
Eq. (\ref{surv_0t.1}). 
A similar calculation, keeping track of ${\cal P}_+(p,s)$ and ${\cal P}_{-}(p,s)$ 
separately
and then inverting the Laplace transform, 
gives
\be
S_+(0,t) - S_-(0,t) = b_+ e^{-\gamma t} ( I_0(\gamma t) - I_1(\gamma t)) 
\label{surv_diff0t.1}
\ee 
where $S_\pm(0,t)$ are the survival probabilities up to time $t$ with {\em final} 
velocity $\pm v_0$ at time $t$,
with $S(0,t) = S_+(0,t) + S_-(0,t)$. 
They satisfy $S_+(0,0)=b_+$ and $S_-(0,0)=b_-$, and $S_+(0,0^+)=b_+$ and $S_-(0,0^+)=0$. 
The ratio of the surviving probabilities is thus 
$S_-(0,t)/S_+(0,t) = I_1(\gamma t)/I_0(\gamma t)$ which is $\simeq \frac{\gamma}{2} t$ 
at small
time and $\simeq 1 - \frac{1}{2 \gamma t}$ at large $t$. 
This is consistent with an equilibration
between the two states at large time and far from the wall. 

The analogue of Eq. (\ref{Pxt_lt.1}), in the inhomogeneous case is
\begin{equation}
\int_0^{\infty} P(x,t|x_0=0)\, e^{-s\,t}\,dt=
\frac{2\,b_+\, \sqrt{s+2\gamma}}{v_0\, \left(\sqrt{s+2\gamma}+\sqrt{s}\right)}\,
e^{-\sqrt{ \frac{s(s+2\,\gamma)}{v_0^2}}\, x}\, .
\label{Pxt_lt.1_inh}
\end{equation}
It turns out that the time reversal symmetry, found in Eq. (\ref{time_reversal.1}) for
the special case $b_{\pm}=1/2$, is no longer valid for generic $b_{+}\ne 1/2$. 

\vskip 0.4cm

{\noindent {\it Late time asymptotic behavior of $S(x_0,t)$.} We conclude this section 
with the following main observation on the late time behavior of the survival
probability $S(x_0,t)$ for generic inhomogeneous initial condition. Clearly, the 
relation $\partial_t S(x_0,t)= - v_0\, P(0,t|x_0)$ in Eq. (\ref{relation.1}) holds for generic 
$b_+$. Substituting the asymptotic large time decay of $P(0,t|x_0)$ from Eq. (\ref{p-0t_asymp.1.inh}) in
this relation then
provides the large $t$ decay of $S(x_0,t)$ for fixed $x_0$ and $b_+$
\begin{equation}
S(x_0,t) \simeq \frac{1}{\sqrt{\pi\, D\, t}}\, 
\left(x_0+ \xi_{\rm Milne}\right)\, ;  \quad\quad{\rm where}\quad D= \frac{v_0^2}{2\gamma}
\label{surv_1p.1_inh}
\end{equation}
and the constant $\xi_{\rm Milne}$ is given exactly by 
\begin{equation}
\xi_{\rm Milne}= b_+\, \frac{v_0}{\gamma}\, .
\label{Milne_1p.1}
\end{equation}
It is instructive to compare our result in Eq. (\ref{surv_1p.1_inh}) with the one for a passive Brownian particle. 
In the latter case, we recall from the introduction that the survival probability $S(x_0,t) \sim x_0/\sqrt{\pi\, D\, t}$
at late times. In the case of the RTP, $S(x_0,t)$ in Eq. (\ref{surv_1p.1_inh}) again decays with same algebraic law $t^{-1/2}$
as in the passive Brownian case with an effective diffusion constant $D=v_0^2/{2\gamma}$, but there is one important and crucial
difference between the two cases. The amplitude $x_0$ of the power-law $t^{-1/2}$ decay in the passive case 
vanishes exactly at $x_0=0$, i.e., if the particle starts
at the wall. In contrast, for the active RTP the amplitude $(x_0+\xi_{\rm Milne})$ approaches a nonzero constant
$\xi_{\rm Milne}= b_+ v_0/\gamma$ as $x_0\to 0$. Thus, even if the RTP starts at the wall, with a finite probability
it can survive up to time $t$. Thus, the late time survival probability for the RTP is exactly of the same form
as in the passive case, but with an effective diffusion constant $D= v_0^2/(2\gamma)$ and
an effective initial distance from the wall $(x_0+ \xi_{\rm Milne})$. In other words, at late times
an active RTP behaves identically to a passive Brownian but with the location of the absorbing
wall effectively shifted from the origin to $-\xi_{\rm Milne}=- b_+ v_0/\gamma$. This effective `extrapolation' length
or `shifting of the wall' also happens in a class of neutron scattering problems where the shift is known
as the Milne extrapolation length---hence we have denoted
it by $\xi_{\rm Milne}$. Similar Milne-like extrapolation lengths also emerge in certain trapping problems of discrete-time
random walks~\cite{MCZ_2006,ZMC_2007,MMS_2017}. Thus our main conclusion from this section is that while the exponent $1/2$
characterizing the power-law decay of $S(x_0,t)$ is the same for both the passive Brownian and the active RTP, the
fingerprint of the `activeness' actually is manifest in the amplitude of this power-law decay (and not in the exponent). While an active RTP
has a nonzero Milne extrapolation length $\xi_{\rm Milne}= b_+ v_0/\gamma >0$, for a passive Brownian motion 
$\xi_{\rm Milne}=0$ identically.}

\section{Two non-crossing RTP's on a line}

{ In this section we consider two independent RTP's on a line and we are interested
in computing the probability that they do not cross each other up to time $t$.
As discussed in the introduction, unlike the Brownian particles, the first-passage
probability for the two-RTP problem can not be reduced to that of a single RTP
in the presence of an absorbing wall at the origin. In this section,
we show that the first-passage probability in this non-Markovian two-RTP problem
can nevertheless be fully solved, using a straightforward generalisation
of our techniques developed in the previous section for a single RTP problem.}   

\begin{figure}
\includegraphics[width=0.8\textwidth]{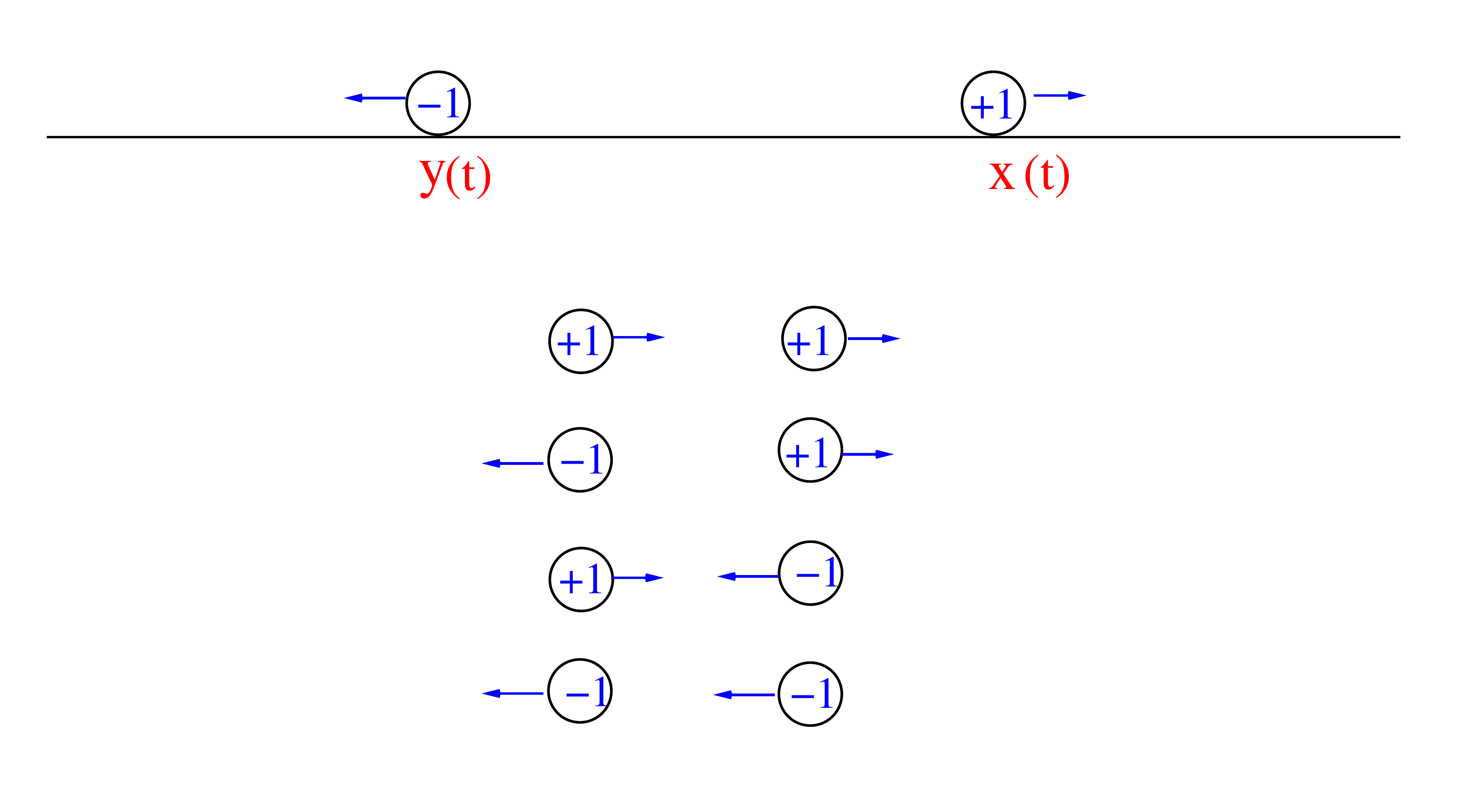}
\caption{Two RTP's on a line. The position of the particle on the right (left) are denoted
respectively by $x(t)$ and $y(t)$, with initial positions $x(0)>y(0)$. 
The internal state $\sigma_1(t)$ and $\sigma_2(t)$ associated with 
the two particles can be in four possible
configurations: $++$, $+-$, $-+$ and $--$, as shown in the figure.}
\label{fig_twortp}
\end{figure}

We consider two RTP's on a line whose
positions $x(t)$ (the particle on the right in Fig. \ref{fig_twortp}) and $y(t)$ 
(the particle on the left) evolve in time independently via the Langevin equations
\begin{equation}
\frac{dx}{dt}= v_0\, \sigma_1(t)\, , \quad \frac{dy}{dt}= v_0\, \sigma_2(t)\, ,
\label{2rtp_lange.1}
\end{equation}
where $\sigma_1(t)$ and $\sigma_2(t)$ are two independent telegraphic noises. For simplicity, we assume
that the intrinsic speed $v_0$, as well as the noise flipping rate $\gamma$ for both particles are the same, 
though our results can be straightforwardly generalised to the cases { when
the parameters of the two noises are different.} The particles
start initially at $x(0)>y(0)$. We are interested in computing the probability that the two particles
do not cross each other up to time $t$ (note that they may encounter each other, but not cross each other). 

We define $P_{\sigma_1,\sigma_2}(x,y,t)$ as the joint probability that (i) 
the right particle reaches $x$ at time $t$ with {internal state} $\sigma_1$ (ii) the left particle
reaches $y$ at time $t$ with {internal state} $\sigma_2$ and (iii) they do not 
{ cross
each other} up to $t$.
There are thus four possibilities denoted respectively by $P_{++}$, $P_{+-}$, $P_{-+}$ and $P_{--}$.
The total probability is obtained by summing over the internal states
\begin{equation}
P(x,y,t)= P_{++}(x,y,t)+ P_{+-}(x,y,t)+ P_{-+}(x,y,t)+ P_{--}(x,y,t)\, .
\label{2total.1}
\end{equation}
Following the method for a single RTP, one can easily write down the Fokker-Planck equations for these
probabilities 
\bea
&& \partial_t P_{++} = - v_0 \partial_x P_{++}  
- v_0 \partial_y P_{++}  - 2 \gamma P_{++}  
+ \gamma (P_{+-}+P_{-+} )\\
&& \partial_t P_{+-} = - v_0 \partial_x P_{+-}  
+ v_0 \partial_y P_{+-}  - 2 \gamma P_{+-}  
+ \gamma (P_{++}+P_{--} ) \\
&& \partial_t P_{-+} = v_0 \partial_x P_{-+}  
- v_0 \partial_y P_{-+}  - 2 \gamma P_{-+}  
+ \gamma (P_{++}+P_{--} ) \\
&& \partial_t P_{--} =  v_0 \partial_x P_{--}  
+ v_0 \partial_y P_{--}  - 2 \gamma P_{--}  
+ \gamma (P_{+-}+P_{-+} ) \;.
\label{2_fp.1}
\eea
We now introduce the non-crossing condition restricting to $x(t)>y(t)$, i.e., the process stops if the two 
particles cross each other. This non-crossing condition can be incorporated via the 
appropriate boundary condition 
\be
P_{+-}(x=y,t) = 0 \;.
\label{2_noncrossing.1}
\ee 
This condition can again be deduced by considering {the} time evolution of a trajectory starting at $x=y$ during
a small interval $dt$, and taking the $dt\to 0$, as in the single RTP case. Indeed observing the state 
where $x(t)$ has velocity $+ v_0$, $y(t)$ has velocity $-v_0$, and $x(t)=y(t)$ necessarily means that the two particles 
have crossed before $t$, which is not allowed. Once again, as we show below,
this single boundary condition at $x=y$, along with the Dirichlet boundary conditions as 
$x\to \infty$ and $y\to - \infty$, are sufficient to uniquely determine the solution to the
Fokker-Planck equations~(\ref{2_fp.1}). We can work with general initial conditions, but for simplicity
we set $x(0)=z_0$ and $y(0)=-z_0$, equidistant from the origin on opposite sides. 
The initial condition is given by 
\be
P_{\sigma_1,\sigma_2}(x,y,t=0) = \left(b_{++},b_{+-},b_{-+},b_{--}\right)\, \delta\left(\frac{x-y}{2} - z_0\right) \delta\left(
\frac{x+y}{2}\right) 
\label{2_init.1}
\ee 
where the $b_{\pm \pm}$ denote the initial probabilities of the $4$ internal state configurations,
with $b_{++}+b_{+-}+b_{-+}+b_{--}=1$.

To solve these equations (\ref{2_fp.1}), it is convenient to go to the center of mass 
and relative coordinates, i.e., we make the change of variables
\bea
w= \frac{x+y}{2} \quad , \quad z= \frac{x-y}{2}  \;.
\label{2_cov.1}
\eea 
In this new pair of coordinates the probability $P_{\sigma_1, \sigma_2}(x,y,t)$
is a different function of $w$ and $z$. But to avoid explosion of new symbols
and with a slight abuse of notations, we will
continue to denote it by $P$, i.e., by $P_{\sigma_1,\sigma_2}(w,z,t)$. Then Eqs. (\ref{2_fp.1}) become
\bea
&& \partial_t P_{++} = - v_0 \partial_w P_{++}  
 - 2 \gamma P_{++}  
+ \gamma (P_{+-}+P_{-+} )\\
&& \partial_t P_{+-} = - v_0 \partial_z P_{+-}  
 - 2 \gamma P_{+-}  
+ \gamma (P_{++}+P_{--} ) \\
&& \partial_t P_{-+} = v_0 \partial_z P_{-+}  
  - 2 \gamma P_{-+}  
+ \gamma (P_{++}+P_{--} ) \\
&& \partial_t P_{--} =  v_0 \partial_w P_{--}  
 - 2 \gamma P_{--}  
+ \gamma (P_{+-}+P_{-+} )
\label{2_fp.2}
\eea
Note that the center of mass $w(t)$ can be any real number (positive or negative), but the relative
coordinate $z(t)>0$ is in the positive half-space, starting from the initial value $z_0>0$.
The boundary condition (\ref{2_noncrossing.1}) now translates into
\begin{equation}
P_{+-}(w, z=0, t)=0\, .
\label{2_zbound.1}
\end{equation}

To proceed, we first define Fourier-Laplace transforms in space
\begin{equation}
{\tilde P}_{\sigma_1,\sigma_2}(k,p,t)= \int_{-\infty}^{\infty} dw\, \int_0^{\infty} dz\,  
e^{-i\, k\, w}\, e^{-p\,z}\, P_{\sigma_1,\sigma_2}(w,z,t)\, .
\label{2_FL.1}
\end{equation} 
Furthermore, we will also take the Laplace transform with respect to time and define
\begin{equation}
{\cal P}_{\sigma_1,\sigma_2}(k,p,s)= \int_0^{\infty} dt\, e^{-s\,t}\, {\tilde P}_{\sigma_1,\sigma_2}(k,p,t)\, .
\label{2_FLL.1}
\end{equation}
Taking these Fourier-Laplace transforms of Eq. (\ref{2_fp.2}) and using the boundary condition
(\ref{2_zbound.1}) we get
\bea
&& s {\cal P}_{++} - {\tilde P}_{++}(k,p,t=0) = i\,v_0 \,k\,  {\cal P}_{++}  
 - 2 \gamma {\cal P}_{++}  
+ \gamma ({\cal P}_{+-}+{\cal P}_{-+} )\\
&& s {\cal P}_{+-} - {\tilde P}_{+-}(k,p,t=0) = - v_0\, p\, {\cal P}_{+-} 
 - 2 \gamma {\cal P}_{+-}  
+ \gamma ({\cal P}_{++}+{\cal P}_{--} ) \\
&& s {\cal P}_{-+} - {\tilde P}_{-+}(k,p,t=0) = v_0 p {\cal P}_{-+}  - v_0\, q_{-+}(k,0,s)
  - 2 \gamma {\cal P}_{-+}  
+ \gamma ({\cal P}_{++}+ {\cal P}_{--} ) \\
&& s {\cal P}_{--} - {\tilde P}_{--}(k,p,t=0) =  -i\, v_0\, k\,  {\cal P}_{--}  
 - 2 \gamma {\cal P}_{--}  
+ \gamma ({\cal P}_{+-}+ {\cal P}_{-+} )\;,
\label{2_fp.3}
\eea
where we have defined
\begin{equation}
q_{-+}(k,0,s)= \int_0^{\infty} dt\, e^{-s\,t}\,\int_{-\infty}^{\infty} dw\, e^{-i\,k\,w}\, P_{-+}(w,z=0,t)\,
\label{2_unknown.1}
\end{equation}
which still remains unknown and will be self-consistently determined using the pole-cancelling mechanism
as in the single RTP case. Note that the initial condition in Eq. (\ref{2_init.1}) implies, putting
$t=0$ in Eq. (\ref{2_FL.1}),
\begin{equation}
{\tilde P}_{\sigma_1,\sigma_2}(k,p,t=0)= e^{-p\, z_0} \, \left(b_{++},b_{+-},b_{-+},b_{--}\right)\, .
\label{2_init.2}
\end{equation}  

Substituting the initial condition (\ref{2_init.2}) on the left hand side (lhs) of Eq. (\ref{2_fp.3})
and inverting the $4\times 4$ matrix gives 
\bea
&& {\cal P}_{\sigma_1,\sigma_2}(k,p,s)  \\
&& = \begin{pmatrix} 
- v_0 i k + 2 \gamma + s & - \gamma & - \gamma & 0 \\
-  \gamma & v_0 p + 2 \gamma + s & 0 &- \gamma \\
 - \gamma & 0 & - v_0 p + 2 \gamma + s & - \gamma \\
 0 &-  \gamma & - \gamma & v_0 i k + 2 \gamma + s 
 \end{pmatrix}^{-1} 
 \left(
\begin{pmatrix}
0 \\
0 \\
- v_0 q_{-+}(k,z=0,s) \\
0
\end{pmatrix}
+ e^{- p z_0} 
\begin{pmatrix}
b_{++} \\
b_{+-}  \\
b_{-+}  \\
b_{--} 
\end{pmatrix}
 \right) \;. \nonumber 
\label{2_invert.1}
\eea 
{ After inverting the $4\times 4$ matrix using Mathematica, we obtain ${\cal 
P}_{\sigma_1,\sigma_2}(k,p,s)$ explicitly. The resulting expressions
are too long to display and are not very illuminating. Summing over
the internal states, the Fourier-Laplace transform of the 
total probability density is given by
\begin{equation}
{\cal P}(k,p,s)= {\cal P}_{++}(k,p,s)+ {\cal P}_{+-}(k,p,s)+ {\cal P}_{-+}(k,p,s)+ {\cal P}_{--}(k,p,s)\, .
\label{2_totalP.1}
\end{equation}
But even this expression is too long for arbitrary initial conditions. Hence
we just present the result for the fully symmetric 
case $b_{++}=b_{+-}=b_{-+}=b_{--} = \frac{1}{4}$ which is a bit simpler, and restore the 
general $b_{\sigma_1,\sigma_2}$ in some of the final results.

For this symmetric initial condition $b_{++}=b_{+-}=b_{-+}=b_{--} = \frac{1}{4}$, we get}
\bea
&& {\cal P}(k,p,s) \label{2_calP_total.1}\\
&& = \frac{e^{-p z_0} \left((2 \gamma +s) \left(v_0^2 (k-p) (k+p)+2 (2 \gamma +s) (4 \gamma
   +s)\right)-2 v_0 e^{p z_0} q_{-+}(k,0,s) \left(k^2 v_0^2+(2 \gamma +s) (4 \gamma
   +s)\right) \left(2 \gamma +p v_0+s\right)\right)}{2 \left(-k^2 p^2 v_0^4+v_0^2 (k-p)
   (k+p) (2 \gamma +s)^2+s (2 \gamma +s)^2 (4 \gamma +s)\right)} \nonumber 
\eea
To fix the unknown $q_{-+}(k,0,s)$, we look for the poles of the rhs of Eq. (\ref{2_calP_total.1}) 
in the complex $p$ plane.
They are located~at
\bea
p^*_{\pm} = \pm \frac{(2 \gamma +s) \sqrt{k^2 v_0^2+s^2+4 \gamma  s}}{v_0 \sqrt{k^2 v_0^2+(2
   \gamma +s)^2}} \;.
\label{2_poles.1}
\eea 
Using the pole-cancelling argument as in the previous section, 
the numerator of the rhs in Eq. (\ref{2_calP_total.1}) must vanish
at the positive pole $p^*_{+}$. This leads to a long but explicit formula for
the unknown $q_{-+}(k,0,s)$
\bea
q_{-+}(k,0,s) = \frac{\left(k^2 v_0^2+(2 \gamma +s) (4 \gamma +s)\right) \exp \left(-\frac{z_0 (2 \gamma
   +s) \sqrt{k^2 v_0^2+s (4 \gamma +s)}}{v_0 \sqrt{k^2 v_0^2+(2 \gamma +s)^2}}\right)}{2
   v_0 \left(\sqrt{k^2 v_0^2+(2 \gamma +s)^2} \sqrt{k^2 v_0^2+s (4 \gamma +s)}+k^2
   v_0^2+(2 \gamma +s)^2\right)} \;.
\label{2_qfixed.1}
\eea
{ A similar but more complicated expression for $q_{-+}(k,0,s)$ can be
obtained explicitly for the inhomogeneous initial condition (with
arbitrary $b_{\sigma_1,\sigma_2}$), but we do not display it here.}
Note from the definition (\ref{2_unknown.1}) that $q_{-+}(k,0,s)$ is just the Fourier-Laplace transform of 
$P_{-+}(w, z=0, t)$. 
{In addition, if we set $k=0$ in Eq. (\ref{2_unknown.1}), i.e., we
integrate over the center of mass coordinate $w$, we get
\begin{equation}
q_{-+}(0,0,s)= \int_0^{\infty} dt\, e^{-s\,t}\, \int_{-\infty}^{\infty} dw\, P_{-+}(w,z=0,t)\, .
\label{q-+00s.1}
\end{equation}
The quantity $\int_{-\infty}^{\infty} dw\, P_{-+}(w,z=0,t)$ has the dimension of the inverse length 
and is proportional to the probability of `reaction' or `encounter' of the two particles in the state $(-+)$ at time $t$ without crossing
each other for all $0 \leq t'<t$, starting at an initial separation $2z_0$. In fact, one can define an encounter probability 
density at time $t$ (with the dimension of the inverse time) as
\begin{equation}
p_{\rm enc}(t|z_0)= v_0\, \int_{-\infty}^{\infty} dw\, P_{-+}(w,z=0,t)\, .
\label{def_enc}
\end{equation}
This nonzero `encountering' probability is a typical hallmark of active particles---it strictly vanishes for
passive Brownian particles. Our analysis thus gives access to this 
nontrivial encountering probability.
Setting $k=0$ in Eq. (\ref{2_qfixed.1}), { or more generally in the
counterpart of Eq. (\ref{2_qfixed.1}) for arbitrary $b_{\sigma_1,\sigma_2}$}, we get 
({ thus restoring the 
dependence on the initial probabilities})
\begin{equation}
\int_0^{\infty} p_{\rm enc}(t|x_0)\, e^{-st}\, dt=
v_0 \, q_{-+}(0,0,s) = \frac{(2 \gamma + (b_{-+}+b_{+-}) s 
+  (b_{-+}-b_{+-}) \sqrt{s(4 \gamma +s)}) e^{-\frac{\sqrt{s(4 \gamma +s)}}{v_0}\, z_0}}{  \left(s + 2
\gamma +\sqrt{s} \sqrt{4 \gamma +s}\right)}\, .
\label{2_encounter.1}
\end{equation}

Note that $\int_0^{\infty} p_{\rm enc}(t|z_0)\, dt= 1$ and hence $p_{\rm enc}(t|z_0)$ has the interpretation
of a probability density of encounter between time $t$ and $t+dt$, starting from $z_0$. 
Remarkably, this Laplace transform can be inverted exactly for all $t$ (see Appendix B). The solution can be conveniently
expressed at all times $t$ in a scaling form
\begin{equation}
p_{\rm enc}(t|z_0)= \gamma\, G\left(\frac{2\gamma\,z_0}{v_0}\, , 2\,\gamma\,t\right)
\label{enc_scaling}
\end{equation}
where the scaling function $G(y,T)$ is given exactly by 
\bea
&& G(y,T)=2 e^{-T}\bigg[b_{-+} \,  \delta(T-y) + \frac{1}{y+T}\left(2 b_{+-}\frac{T-y}{T+y}+(b_{++}+ b_{--})y\right)\,I_0(\rho) 
\nonumber \\
&& +\frac{1}{\rho}\left(y(b_{-+}+ b_{+-}) + (b_{++}+ b_{--}) \frac{T-y}{T+y}- 2 b_{+-} \frac{1}{T+y}\left(y^2+\frac{2(T-y)}{T+y}\right) \right)\, I_1(\rho)\bigg]
\theta(T-y)  \;,
\label{enc_exact}
\eea
where $\rho= \sqrt{T^2-y^2}$. In Fig. \ref{Fig_enc} we show a plot of $p_{\rm enc}(t|z_0)$, given in Eqs. (\ref{enc_scaling}) and (\ref{enc_exact}), for $b_{\pm \pm}=1/4$, as a function of $t$ and for two different values of $z_0$. {Note that $p_{\rm enc}(t|z_0) = 0$ for $t < z_0/v_0$ since $z_0/v_0$ is the minimal time needed for the two particles to encounter (this corresponds to pairs ($-+$) whose velocities have not changed up to that time). The limiting behavior of $p_{\rm enc}(t|z_0)$ when $t \to (z_0/v_0)^+$ can be obtained from the explicit expression (\ref{enc_exact}). It has both a singular part $\propto \delta(t-z_0/v_0)$ as well as a regular finite part (see Fig. \ref{Fig_enc}) and reads
\bea\label{smallt_enc}
p_{\rm enc}(t|z_0) \to b_{-+} \, e^{-\frac{2\gamma z_0}{v_0}} \delta\left(t - \frac{z_0}{v_0}\right) + 2 \gamma e^{-\frac{2\gamma z_0}{v_0}}  \left( \frac{b_{++} + b_{--}}{2} + \frac{\gamma z_0}{v_0} b_{-+} \right) \;, \;\;\;  t \to (z_0/v_0)^+ \;.
\eea
}   

\begin{figure}
\includegraphics[width = 0.65\linewidth]{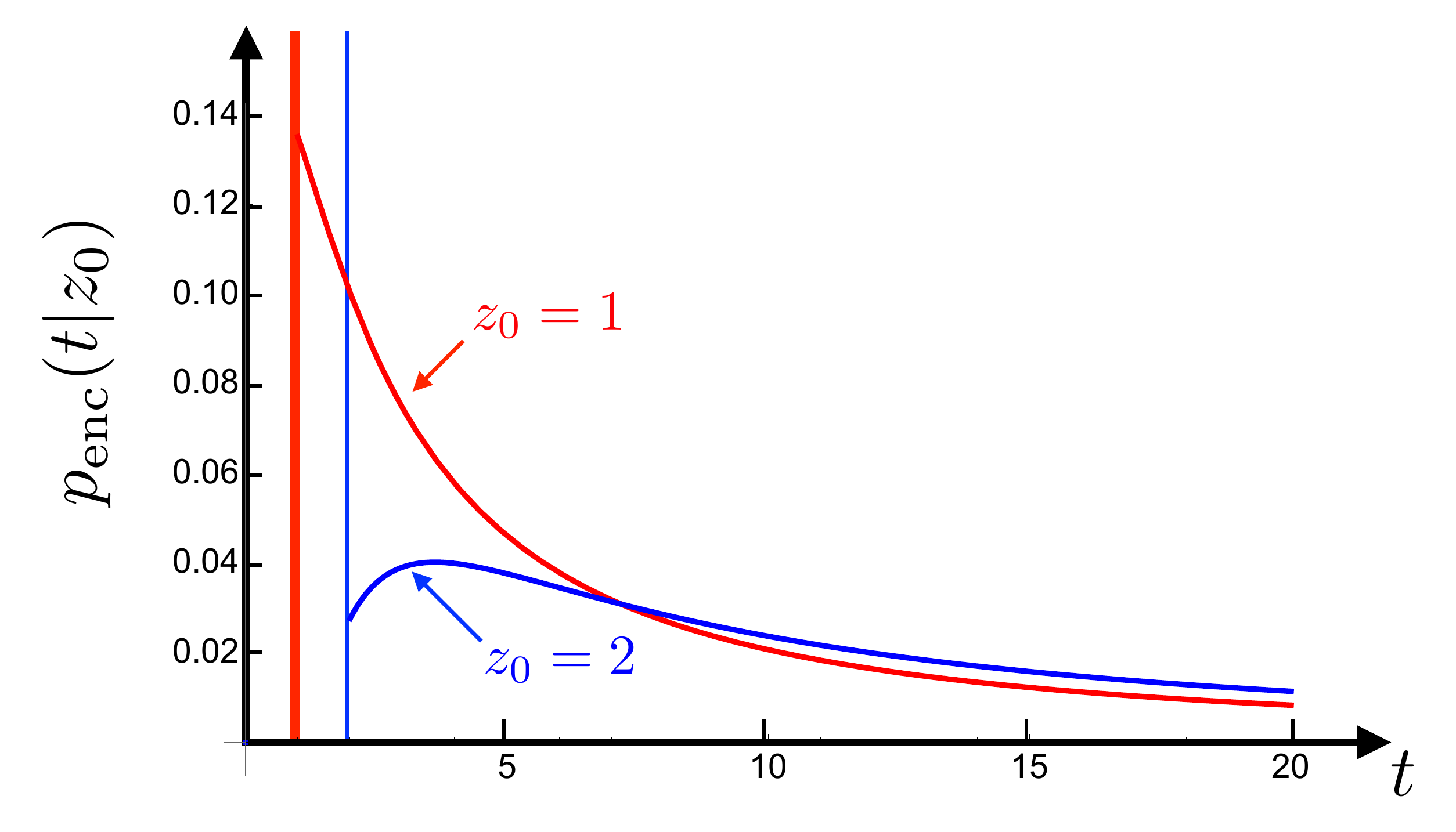}
\caption{{Plot of $p_{\rm enc}(t|z_0)$, as given in Eqs. (\ref{enc_scaling}) and (\ref{enc_exact}) as a function of $t$ and for two different values of $z_0 = 1$ (red) and $z_0 = 2$ (blue) with parameter values $v_0=1$ and $\gamma=1$ and symmetric initial conditions $b_{\pm \pm} = 1/4$. The vertical colored lines correspond to the delta-function in Eqs. (\ref{enc_exact}) and (\ref{smallt_enc}) at $t=z_0$, whose weight decreases exponentially with $z_0$. These correspond to pairs $(-+)$ which have not changed their
velocities up to time $t = z_0/v_0$.}}\label{Fig_enc}
\end{figure}

The large $t$ behavior of $p_{\rm enc}(t|z_0)$ for fixed $z_0$ can be easily obtained by
analysing the small $s$ behavior of $q_{-+}(0,0,s)$. Taking the small $s$ limit on the rhs of
Eq. (\ref{2_encounter.1}) we get
\begin{equation}
v_0 \, q_{-+}(0,0,s)= 1- \frac{1}{\sqrt{\gamma}}\,\left(1 + b_{+-} - b_{-+} + \frac{2\gamma}{v_0}\, z_0
\right)\, \sqrt{s} +O(s)\, .
\label{2_encounter_smalls.1}
\end{equation}
Consequently, upon inverting {and} using a Tauberian theorem, we find that the encountering probability
at late times decays algebraically as
\begin{equation}
p_{\rm enc}(t|z_0) \approx \frac{1}{\sqrt{{4\pi \gamma}}}\left(1 + b_{+-} - b_{-+}  + \frac{2\gamma}{v_0}\, z_0\right)\,
\frac{1}{t^{3/2}}\, .
\label{p_enc_asymp.1}
\end{equation}
The same result also follows from the exact form in Eq. (\ref{enc_exact}).
}
Extending the calculation to obtain the encounter probabilities associated to
the pairs $(++)$ and $(--)$, i.e. $\int_{-\infty}^{+\infty} dw \, P_{++}(w,z=0,t)$ and 
$\int^{+\infty}_{-\infty} dw \, P_{--}(w,z=0,t)$, we find that at large time they both decay as $t^{-3/2}$
with the same amplitude as $p_{\rm enc}(t|z_0)$ up to a factor $1/2$, i.e. both quantities are equivalent to  
$\frac{1}{2} p_{\rm enc}(t|z_0)$ for large time $t$.

Substituting the exact $q_{-+}(k,0,s)$ from Eq. (\ref{2_qfixed.1}) into (\ref{2_calP_total.1}) finally
gives  (for $b_{\pm \pm}=1/4$)
\bea
&& {\cal P}(k,p,s) = \\
&&
\frac{e^{-p z_0} \left((2 \gamma +s) \left(v_0^2 (k-p) (k+p)+2 (2 \gamma +s) (4 \gamma
   +s)\right)-\frac{\left(k^2 v_0^2+(2 \gamma +s) (4 \gamma +s)\right){}^2 \left(2 \gamma
   +p v_0+s\right) \exp \left(z_0 \left(p-\frac{(2 \gamma +s) \sqrt{k^2 v_0^2+s^2+4
   \gamma  s}}{v_0 \sqrt{k^2 v_0^2+(2 \gamma +s)^2}}\right)\right)}{\sqrt{k^2 v_0^2+s^2+4
   \gamma  s} \sqrt{k^2 v_0^2+(2 \gamma +s)^2}+k^2 v_0^2+(2 \gamma +s)^2}\right)}{2
   \left(-k^2 p^2 v_0^4+v_0^2 (k-p) (k+p) (2 \gamma +s)^2+s (2 \gamma +s)^2 (4 \gamma
   +s)\right)} \nonumber
\label{2_calP_total.2}
\eea 
This rather long (albeit explicit) expression simplifies a bit by setting $k=0$, i.e., integrating
over the center of mass coordinate 
\begin{equation}
{\cal P}(k=0,p,s) = 
\frac{e^{-p z_0}}{2 (2 \gamma +s) \left(-p^2 v_0^2+s^2+4\,\gamma\, s\right)}\,  
\left[-p^2 v_0^2 +
2\, (2 \gamma +s)\, (4 \gamma +s)
-\frac{(2\gamma+v_0\, p+s)(4\gamma+s)^2}{2\gamma +s+\sqrt{s(s+4\gamma)}}\, 
e^{z_0\,\left(p-\frac{\sqrt{s(s+4\gamma)}}{v_0}\right) }\right]\, .
\label{2_calP_total.3}
\end{equation}
It behaves as $e^{-p z_0}/s$ at large $s$, consistent with the initial condition. From this exact formula (\ref{2_calP_total.3}) one
can also check that $\int_{-\infty}^{+\infty} dw \, P(w,z=0,t) \simeq 2 \,p_{\rm enc}(t|z_0)$, for $t \gg 1$, which is fully
consistent with our previous results mentioned below Eq. (\ref{p_enc_asymp.1}). 

Finally, the survival probability $S(z_0,t)$, i.e., the probability that the two particles, starting
{initially at a separation $z_0$, does not cross each other up to time $t$ is obtained by integrating
over all $z$, i.e., by setting $p=0$ in Eq. (\ref{2_calP_total.3}). We get (restoring the dependence
in the initial probabilities)
{ \begin{equation}
\int_0^{\infty} S(z_0,t)\, e^{-s\,t}\,dt = {\cal P}(k=0,p=0,s) 
 = \frac{1}{s}\left[1- \frac{(4 \gamma +2 (b_{-+}+b_{+-}) s 
+ 2 (b_{-+}-b_{+-}) \sqrt{s(4 \gamma +s)})
}{2\,(2\gamma+s+ \sqrt{s(s+4\gamma)} )
 }\, 
e^{-\frac{\sqrt{s(s+4\gamma)}}{v_0}\, z_0}\right]\, .
\label{2_surv.1}
\end{equation} }
Interestingly, by comparing this relation (\ref{2_surv.1}) with the result obtained above for the encounter probability $p_{\rm enc}(t|z_0)$ in Eq. (\ref{2_encounter.1}), we find the following identity
\begin{equation}\label{relation.2}
\partial_t S(z_0,t) = - p_{\rm enc}(t |z_0) \;,
\end{equation}
which is analogous to the identity found in Eq. (\ref{relation.1}) for the case of a single particle with an absorbing wall at the origin. As above [see the discussion below Eq. (\ref{relation.1})], \eqref{relation.2} can be 
obtained by summing all four equations in \eqref{2_fp.2} and integrating for $w \in ]-\infty,\infty[$ and
$z \in [0,+\infty[$.
This identity clearly shows that $p_{\rm enc}(t |z_0) dt$ is the probability that the
two particles encounter in the state $(-+)$, {\it and hence die immediately}, in the
time interval $[t,t+dt[$. It is thus the first and last encounter of the two particles in the
state $(-+)$. {Again, we emphasize that this relation (\ref{relation.2}) is a specific feature of active particles, which does not hold for passive (i.e. Brownian) ones. Indeed, for Brownian particles, the encounter probability is strictly zero, while the first-passage probability is not, since the probability current at $z=0$ is non-zero.}

The relation \eqref{relation.2}, together with the scaling form for the encounter probability (\ref{enc_scaling}),
leads to the following explicit result for the survival probability
\begin{equation}
S(z_0,t)  = H\left(\frac{2\gamma\,z_0}{v_0}\, , 2\,\gamma\,t\right) \quad , \quad 
\partial_T H(y,T) = - \frac{1}{2} G(y,T) 
\label{enc_scalingS}
\end{equation}
where $G(y,T)$ is given explicitly in \eqref{enc_exact}. {In the special case $z_0=0$ we obtain explicitly
\bea \label{S0t_exact}
S(0,t) = e^{-2 \gamma t}\left( (1+ b_{+-}-b_{-+})\left(I_0(2\gamma t) + I_1(2 \gamma t) \right)   - \frac{b_{+-}}{\gamma t} I_1(2 \gamma t)\right) \;, \; t >0 \;.
\eea
Its asymptotic behaviors are easily obtained as $S(0,t) \to (1-b_{-+})$ for $t \to 0^+$, as expected since between $t=0$ and $t=0^+$ the pairs $-+$ necessarily annihilate, while $S(0,t) \approx (1+b_{+-}-b_{-+})/\sqrt{\pi \gamma t}$ for $t \to \infty$.
}

{For arbitrary $z_0 >0$, one can easily extract the late time behavior of $S(z_0,t)$ from the Laplace transform in Eq.~(\ref{2_surv.1}).} 
Indeed, expanding for small $s$ gives
\bea
{\cal P}(k=0,p=0,s) = \frac{v_0(1 + b_{+-} - b_{-+}) +2 \gamma  z_0}{\sqrt{\gamma } \sqrt{s} v_0} +O(1) \;.
\label{2_surv_smalls.1}
\eea
Inverting we obtain the large time decay of the no-crossing probability
\bea
S(z_0, t) \simeq \frac{v_0(1 + b_{+-} - b_{-+}) +2 \gamma  z_0}{\sqrt{\pi\, v_0^2\, \gamma\, t}}\, ,
\label{2_surv.2}
\eea 
which is consistent with the result obtained above for $z_0=0$ in Eq. (\ref{S0t_exact}).  {Let us rewrite Eq. (\ref{2_surv.2}) as
\begin{equation}
S(z_0,t) \simeq \frac{1}{\sqrt{\pi\, D'\, t}}\left(z_0+\xi_{\rm Milne}\right)\, ; \quad\quad {\rm where}\quad D'=\frac{v_0^2}{4\gamma}\, ,
\label{2_surv_rep.2}
\end{equation}
and the Milne extrapolation length $\xi_{\rm Milne}$ for this two RTP problem is given by
\begin{equation}
\xi_{\rm Milne}= \frac{v_0}{2\gamma}\,(1+ b_{+-}-b_{-+})\, .
\label{2_Milne.1}
\end{equation}  

Thus the survival probability (i.e., the probability of no crossing of the two independent RTP's) decays as $t^{-1/2}$ at late times, 
as in the case of two independent `passive' Brownian particles.
However, the amplitude of the decay carries an interesting feature, as in the case of a single RTP in the presence of a wall.
As discussed in the introduction, for two independent passive Brownian motions
starting at an initial sepration $2z_0$, the probability of no zero crossing up to time $t$ decays at late times 
as $\sim z_0/\sqrt{\pi\, D'\, t}$ where $D'=D/2$ (see Eq. (\ref{Brown_surv.1})).
Thus, if $z_0=0$, the Brownian particles cross immediately. Hence the amplitude of the $t^{-1/2}$ decay
vanishes at the absorbing boundary. In contrast, we see from Eq. (\ref{2_surv_rep.2}) that in the active case,
the amplitude $z_0+ \xi_{\rm Milne}$ does not vanish when $z_0=0$. This is because even if the two particles
start at the same initial position, with a finite probability they can go away from each other in the
opposite direction and hence survive without crossing each other. The dependence of this amplitude in
the initial probabilities can be understood qualitatively: (i) changing $b_{++}$ or $b_{--}$ 
only affects the motion of the center of mass, hence these probabilities do not appear in
the survival probability (ii) to survive till late times it is clearly advantageous to start in the configuration $(+-)$ rather than 
in $(-+)$. In fact, the amplitude of the late time decay
vanishes, when extrapolated to the negative side, at $z_0=- v_0(1 + b_{+-} - b_{-+}) /{2\gamma}=-\xi_{\rm Milne}$, as in
the case of a single RTP in the presence of a wall.
Clearly, in the passive limit $\gamma\to \infty$, the Milne extrapolation length vanishes. Hence, for two active 
particles also, a finite
Milne extrapolation length is a clear signature of `activeness' of the RTP's dynamics.} 
 
Finally, {if we take the scaling (diffusive) limit corresponding to $s \to 0$, $z_0 \to \infty$ but keeping $\sqrt{s} z_0$ fixed, one finds}
\bea
{\cal P}(k=0,p=0,s) \simeq \frac{1-e^{-\frac{2 \sqrt{\gamma } \sqrt{s} z_0}{v_0}}}{s} \;.
\eea 
{This Laplace transform can be easily inverted to obtain, in real time, }
\bea
S(z_0,t) \simeq
\text{erf}\left(\frac{\sqrt{\gamma } z_0}{\sqrt{t} v_0}\right) \;,
\eea
which is the survival probability of a Brownian walker with diffusion constant $D'=v_0^2/(4 \gamma)$. 
Alternatively, one can keep $z_0$ fixed, but take the limit $v_0\to \infty$, $\gamma\to \infty$
with the ratio $D'= v_0^2/{(4\gamma)}$ fixed. In this case, once again we recover 
the passive Brownian behavior as expected.

%
%
\section{Conclusion}

In this paper we have studied non crossing probabilities for active particles,
in the framework of a simple run and tumble model with velocities $\pm v_0$ 
subjected to a telegraphic noise. We have computed explicitly
the probability of non-crossing of two active RTP's up to time $t$.

We found useful to first consider the case
of a single particle with an absorbing wall. For that problem we have calculated 
explicitly the total probability density $P(x,t|x_0)$ that the particle, starting at $x_0$ at time $t=0$, survives up 
to time $t$ and that it is at position $x$ at $t$. Contrarily to the passive Brownian particle (which is recovered for
$v_0 \sim \sqrt{\gamma} \to +\infty$) the probability of presence at the wall does not vanish and we found that it
decreases as $t^{-3/2}$. Integration of $P(x,t|x_0)$ over $x$ then allows to recover the
survival probability $S(x_0,t)$ obtained previously using different methods in \cite{Malakar_2018,EM_2018}.
Here we showed an interesting exact relation for active RTP dynamics: the probability to find the particle at the wall is (minus) the time derivative
of the survival probability. The latter decays at large time as $t^{-1/2}$. The amplitude of the decay of the survival probability 
explicitly depends on
$x_0$ and $b_+$, where $b_{\pm}$ are the probabilities $b_{\pm}$ that the particle is
in states $\pm v_0$ at time $t=0$. This defines a length scale analogous to the so called ``Milne length'', known from
the neutron-scattering literature. 

We then studied the case of two indepedent RTP's and computed the probability that they do not cross each other up to time $t$.
In the case of
two passive Brownian particles this problem can be mapped exactly to the one of the 
single particle with an absorbing wall. For the active problem this equivalence fails.
By considering all four states for the two particle systems we obtain the double
Laplace transform of the probability that the two particles, initially separated by a distance
$2 z_0>0$, have survived up to time $t$ and are at a distance $2 z$ from each other at time $t$.
From it we have extracted the "encounter" probability, i.e., the
probability that the two particles are at the same position at time $t$. 
At variance with the passive (Brownian) case it is non zero. It decays at large
time as $t^{-3/2}$ with an amplitude which depends on $z_0$ and
on the probabilities of the velocities in the initial state. Similarly to
the absorbing wall problem, the encounter probability is the time derivative
of the survival probability. The survival probability thus again decays at large time as 
$t^{-1/2}$
with an amplitude that is proportional to $(z_0+\xi_{\rm Milne})$. This amplitude thus
vanishes when the initial $z_0$ is extrapolated to the negative side at $z_0=-\xi_{\rm Milne}$.  
We have computed exactly $\xi_{\rm Milne}$ for this two RTP problem.
Our main conclusion is that the amplitude of the $t^{-1/2}$ decay of the late time survival probability
carries a fingerprint of the activeness of the particles: active particles have a finite
Milne extrapolation length $\xi_{\rm Milne}$, while the passive ones have $\xi_{\rm Milne}=0$.

{In this work, we have considered the case of two ``free'' annihilating RTP's. A natural question
is to understand what happens if instead these particles are confined by an external potential, a situation
that has recently attracted much attention for active particles \cite{Dhar_18,Basu_18, Sevilla_19,Dhar_2019}. Another natural
question is whether there exists extensions of the so-called Karlin-McGregor formula \cite{KMG}, valid for passive Brownian particles, 
which would allow to study an arbitrary number of non-crossing RTPs on the line. This is left for future investigations.} 

\acknowledgements

We thank A. Dhar, A. Kundu and S. Sabhapandit for useful discussions. We acknowledge support from ANR grant ANR-17-CE30-0027-01 RaMaTraF.

\appendix
\section{Laplace inversion of Eq. (\ref{ptotal0t.1})}

To invert the Laplace transform in Eq. (\ref{ptotal0t.1}), it is useful to 
re-write the rhs of
Eq. (\ref{ptotal0t.1}) as follows
\begin{equation}
\int_0^{\infty} P(0,t|x_0)\, e^{-s\,t}\, dt  = 
\frac{s+2\gamma-\sqrt{s(s+2\gamma)}}{2v_0\gamma}\, 
e^{-\sqrt{ \frac{s(s+2\,\gamma)}{v_0^2}}\, x_0}\, .
\label{lt_A1}
\end{equation}
In order to bring it to a more amenable form, it is convenient to rescale $t\to t/\gamma$ and 
$s\to \gamma\, s$
and re-express Eq.~(\ref{lt_A1})~as
\begin{equation}
\int_0^{\infty} P\left(0,\frac{t}{\gamma}\big|x_0\right)\, e^{-s\,t}\, dt  =
\frac{\gamma}{2v_0}\,
\left[s+2-\sqrt{s(s+2)}\right]\,
e^{-\sqrt{s(s+2)}\, z}\, ; \quad {\rm where}\,\, z= \frac{\gamma x_0}{v_0} \;.
\label{lt_A2}
\end{equation}
We denote by ${\cal L}_s^{-1}$ the inverse Laplace transform with respect to $s$. Then we invert
Eq. (\ref{lt_A2}) and split the rhs into two separate terms
\begin{equation}
 P\left(0,\frac{t}{\gamma}\big|x_0\right)= 
\frac{\gamma}{2v_0}\, \left(
{\cal L}_s^{-1}\left[
\left(s+1-\sqrt{s(s+2)}\right)\, e^{-\sqrt{s(s+2)}\,z}\right]+
{\cal L}_s^{-1}\left[e^{-\sqrt{s(s+2)}\,z}\right]\right)\, .
\label{ilt_A3}
\end{equation}

The reason behind the splitting of the rhs into two terms is as follows. The Laplace inversion
of the second term is known explicitly (see e.g. Ref.~\cite{Malakar_2018})
\begin{equation}
{\cal L}_s^{-1}\left[e^{-\sqrt{s(s+2)}z}\right]= 
\frac{z\, e^{-t}}{\sqrt{t^2-z^2}}\, I_1\left(\sqrt{t^2-z^2}\right)\ \theta(t-z) + e^{-t}\, \delta(t-z)\, ,
\label{ilt2_A4}
\end{equation}
where $I_1(x)$ is the modified Bessel function and $\theta(x)$ is the standard Heaviside theta function.
The inversion of the first term in Eq. (\ref{ilt_A3}) requires a bit more work. To proceed, we make use
of another interesting Laplace inversion that was found in Ref.~\cite{Malakar_2018}
\begin{equation}
{\cal L}_s^{-1}\left[ \frac{s+1-\sqrt{s(s+2)}}{\sqrt{s(s+2)}}\, e^{-\sqrt{s(s+2)}\,z}\right]
= e^{-t}\, \sqrt{ \frac{t-z}{t+z} }\, I_1\left(\sqrt{t^2-z^2}\right)\, \theta(t-z)\, .
\label{ilt1_A5}
\end{equation}
We then take the derivative with respect to $z$ in Eq. (\ref{ilt1_A5}) and use the identity 
satisfied by the Bessel function: $x\, dI_1(x)/dx + I_1(x)= x\, I_0(x)$. After a few steps
of straightforward algebra we get 
\begin{equation}
{\cal L}_s^{-1}\left[
\left(s+1-\sqrt{s(s+2)}\right)\, e^{-\sqrt{s(s+2)}\,z}\right]= \frac{e^{-t}}{t+z}\left[
z\, I_0\left(\sqrt{t^2-z^2}\right)+ \sqrt{\frac{t-z}{t+z}}\, 
I_1\left(\sqrt{t^2-z^2}\right)\right]\,\theta(t-z)\, .
\label{ilt1_A6}
\end{equation}
Adding Eqs. (\ref{ilt2_A4}) and (\ref{ilt1_A6}) on the rhs of Eq. (\ref{ilt_A3}) and substituting
$t/\gamma\to t$ and $z= \gamma x_0/v_0$, we obtain the result in Eq. (\ref{exact_inv.1}).

\section{Laplace inversion of Eq. (\ref{2_encounter.1})}
{
To invert the Laplace transform in Eq. (\ref{2_encounter.1}), we first make a change of variables
$t\to t'/{2 \gamma}$ and $s\to 2\gamma\, s$ giving
\begin{equation}
{\int_0^{\infty} p_{\rm enc}\left(\frac{t}{2\gamma}| z_0\right) e^{- s t} \,dt}= 2 \gamma\, 
\frac{ s(b_{-+}+b_{+-}) +1 + (b_{-+}-b_{+-}) \sqrt{s(s+2)} }{s+1+\sqrt{s(s+2)}}\, 
e^{-\sqrt{s(s+2)}\, y}\, , \quad {\rm where} \quad y= \frac{2\gamma z_0}{v_0}\, .
\label{enc_lt.B1}
\end{equation} 
Inverting and expressing $s+2= s+1+\sqrt{s(s+2)}+1-\sqrt{s(s+2)}$, we split the rhs into $3$ terms
\begin{eqnarray}
p_{\rm enc}\left(\frac{t}{2\gamma}| z_0\right)= \; &&2 \gamma\, 
\Bigg( (b_{-+}+b_{+-})
{\cal L}_s^{-1}\left[e^{-\sqrt{s(s+2)}\,y}\right]
+ (1- b_{-+}-b_{+-}){\cal L}_s^{-1}\left[\frac{1}{s+1+\sqrt{s(s+2)}}\, e^{-\sqrt{s(s+2)}\,y}\right]  \\
&&- 2 b_{+-}{\cal L}_s^{-1}\left[\frac{\sqrt{s(s+2)}}{s+1+\sqrt{s(s+2)}}\, e^{-\sqrt{s(s+2)}\,y}\right]
\Bigg)\, .
\label{enc_ilt_B2}
\end{eqnarray}
The first term on the rhs can be inverted explicitly using Eq. (\ref{ilt2_A4}). The second term can be written as
\begin{equation}
{\cal L}_s^{-1}\left[\frac{1}{s+1+\sqrt{s(s+2)}}\, e^{-\sqrt{s(s+2)}\,y}\right]=
{\cal L}_s^{-1}\left[\left(s+1-\sqrt{s(s+2)}\right)\, e^{-\sqrt{s(s+2)}\,y}\right]
\label{enc_T2_B3}
\end{equation}
and subsequently can be inverted explicitly using Eq. (\ref{ilt1_A6}). Finally, the third term in Eq. (\ref{enc_ilt_B2})
is just the derivative with respect to $y$ of the second term. Hence, one can also invert it explicitly
by taking derivative of Eq. (\ref{ilt1_A6}) with respect to $z$ and setting $z=y$. Finally, after summing up the three
contributions and using the Bessel function relations, $dI_0(z)/dz=I_1(z)$ and $dI_1(z)/dz= I_0(z)-I_1(z)/z$, we
arrive at the result in Eqs. (\ref{enc_scaling}) and (\ref{enc_exact}). 

\section{Some useful Laplace inversions}
In this appendix we provide a list of Laplace inversions that are not easy to find in the standard literature
and Mathematica is unable to find them. We believe that these inversions would be useful for future works
on active systems where such Laplace transforms occur frequently.
We define ${\cal L}_s^{-1}$ as the inverse Laplace transform of a function whose argument is denoted by $t$, i.e.,
$s$ is conjugate to $t$. Then the following results are true, and one can easily verify them numerically. Below we assume that $z>0$. 
\vskip 0.4cm

\begin{equation}
{\cal L}_s^{-1}\left[e^{-\sqrt{s(s+2)}\,z}\right]=
\frac{z\, e^{-t}}{\sqrt{t^2-z^2}}\, I_1\left(\sqrt{t^2-z^2}\right)\ \theta(t-z) + e^{-t}\, \delta(t-z)\, ,
\label{inv_C1}
\end{equation}

\begin{equation}
{\cal L}_s^{-1} \left[\frac{1}{\sqrt{s(s+2)} }\, e^{-\sqrt{s(s+2)}\, z}\right]
= e^{-t}\, I_0\left(\sqrt{t^2-z^2}\right)\, \theta(t-z)\, .
\label{inv_C2}
\end{equation}

\begin{equation}
{\cal L}_s^{-1}\left[ \frac{s+1-\sqrt{s(s+2)}}{\sqrt{s(s+2)}}\, e^{-\sqrt{s(s+2)}\,z}\right]
= e^{-t}\, \sqrt{ \frac{t-z}{t+z} }\, I_1\left(\sqrt{t^2-z^2}\right)\, \theta(t-z)\, .
\label{inv_C3}
\end{equation}

\begin{equation}
{\cal L}_s^{-1}\left[
\left(s+1-\sqrt{s(s+2)}\right)\, e^{-\sqrt{s(s+2)}\,z}\right]= 
\frac{e^{-t}}{t+z}\left[
z\, I_0\left(\sqrt{t^2-z^2}\right)+ \sqrt{\frac{t-z}{t+z}}\,
I_1\left(\sqrt{t^2-z^2}\right)\right]\,\theta(t-z)\, .
\label{inv_C4}
\end{equation}

\begin{eqnarray}
&&{\cal L}_s^{-1}\left[ \frac{\sqrt{s(s+2)}}{s+1+\sqrt{s(s+2)}}\, e^{-\sqrt{s(s+2)}\,z}\right] \nonumber \\
&&= \frac{e^{-t}}{t+z}\,\left[\frac{1}{\sqrt{t^2-z^2}}\left(z^2+ \frac{2(t-z)}{t+z}\right)I_1\left(\sqrt{t^2-z^2}\right)
- \frac{t-z}{t+z}I_0\left(\sqrt{t^2-z^2}\right)\right]\theta(t-z) + \frac{1}{2}\, e^{-t}\delta(t-z)\, .
\label{inv_C5}
\end{eqnarray}

\begin{equation}
{\cal L}_s^{-1}\left[ \sqrt{\frac{s+2}{s}}\, e^{-\sqrt{s(s+2)}\, z}\right]  = e^{-t}\left[I_0\left(\sqrt{t^2-z^2}\right)
+ \frac{t}{\sqrt{t^2-z^2}}\, I_1\left(\sqrt{t^2-z^2}\right)\right]  \theta(t-z) +  e^{-t}\delta(t-z) \;. 
\label{inv_C6}
\end{equation}

}


\begin{thebibliography}{26}

\bibitem{Chandra_1943} S. Chandrasekhar, {\em Stochastic problems in physics and astronomy},
Rev. Mod. Phys. {\bf 15}, 1 (1943).

\bibitem{Redner_book} S. Redner, {\em A Guide to First-Passage Processes} (Cambridge University Press, 2001).

\bibitem{SM_review} S. N. Majumdar, {\em Persistence in nonequilibrium systems},
Curr. Sci.  {\bf 77}, 370 (1999).

\bibitem{BF_2005} S. N. Majumdar, {\em Brownian Functionals in Physics and Computer Science}, 
Curr. Sci. {\bf 89}, 2076 (2005).  

\bibitem{Persistence_review} A. J. Bray, S. N. Majumdar, and G. Schehr, {\em Persistence and first-passage 
properties in nonequilibrium systems}, Adv. in Phys. {\bf 62}, 225 (2013).

\bibitem{fp_book_2014} {\em First-Passage Phenomena and Their Applications}, Eds. R. Metzler, G. Oshanin, S. Redner 
(World Scientific, 2014).


\bibitem{Berg_book} H. C. Berg, {\em E. coli in Motion} (Springer, 2014).

\bibitem{TC_2008} J. Tailleur and M. E. Cates, {\em Statistical mechanics of interacting Run-and-Tumble
bacteria}, Phys. Rev. Lett. {\bf 100}, 218103 (2008).

\bibitem{ML_2017} J. Masoliver and K. Lindenberg, {\em Continuous time persistent random walk: a review and 
some generalizations}, Eur. Phys. J B {\bf 90}, 107 (2017).

\bibitem{Weiss_2002} G. H. Weiss, {\em Some applications of persistent random walks and the telegrapher's equation}, Physica A: Statistical Mechanics and its Applications, {bf 311}, 381 (2002).

\bibitem{ADP_2014} L. Angelani, R. Di Lionardo, and M. Paoluzzi, {\em First-passage time of run-and-tumble particles}, 
Euro. J. Phys. E {\bf 37}, 59 (2014).

{\bibitem{A2015} L. Angelani, {\em Run-and-tumble particles, telegrapher's equation and absorption problems with partially reflecting boundaries},
 J. Phys. A: Math. Theor. {\bf 48}, 495003 (2015).} 

\bibitem{Malakar_2018} K. Malakar, V. Jemseena, A. Kundu, K. Vijay Kumar, S. Sabhapandit, S. N.
Majumdar, S. Redner, A. Dhar, {\em Steady state, relaxation and
first-passage properties of a run-and-tumble particle in one-dimension}, J. Stat. Mech. P043215 (2018).

\bibitem{DM_2018} T. Demaerel and C. Maes, {\em Active processes in one dimension}, Phys. Rev. E {\bf 97}, 032604 (2018).

\bibitem{EM_2018} M. R. Evans and S. N. Majumdar,
{\em Run and tumble particle under resetting: a renewal approach},
J. Phys. A: Math. Theor. {\bf 51}, 475003 (2018).

\bibitem{MLW_86} J. Masoliver, K. Lindenberg, and B. J. West, {\em First-passage times for non-Markovian 
processes: Correlated impacts on a free process}, Phys. Rev. A {\bf 34}, 1481 (1986).

\bibitem{Dhar_18} A. Dhar, A. Kundu, S. N. Majumdar, S. Sabhapandit, G. Schehr,
{\em Run-and-tumble particle in one-dimensional confining potential: Steady state, relaxation and 
first passage properties}, arXiv: 1811.03808

\bibitem{SEB_16} A. B. Slowman, M. R. Evans, R. A. Blythe,
{\em Jamming and attraction of interacting run-and-tumble random walkers},
Phys. Rev. Lett. {\bf 116}, 218101 (2016).

\bibitem{SEB_17} A. B. Slowman, M. R. Evans, R. A. Blythe, {\em Exact solution of two interacting 
run-and-tumble random walkers with finite tumble duration}, J. Phys. A: Math, Theor.  {\bf 50}, 375601 (2017).

\bibitem{Mallmin_18} E. Mallmin, R. A. Blythe, M. R. Evans, {\em Exact spectral solution of two interacting
run-and-tumble particles on a ring}, arXiv: 1810.00813 

\bibitem{RM_00} R. Rajesh, and S. N. Majumdar, {\em Conserved Mass Models and Particle Systems
in One Dimension}, J. Stat. Phys. {\bf 99}, 943 (2000).

\bibitem{RM_00.1} R. Rajesh, and S. N. Majumdar, {\em Exact Calculation of the Spatio-temporal
Correlations in the Takayasu model and in the q-model
of Force Fluctuations in Bead Packs}, Phys. Rev. E, {\bf 62}, 3186 (2000). 

\bibitem{MCZ_2006} S.N. Majumdar, A. Comtet, and R.M. Ziff,  {\em Unified Solution of the Expected 
Maximum of a Random
Walk and the Discrete Flux to a Spherical Trap}, J. Stat. Phys. {\bf 122}, 833 (2006).

\bibitem{ZMC_2007} R.M. Ziff, S.N. Majumdar and A. Comtet, {\em General Flux to Trap in One and Three
Dimensions}, J. Phys. C: Cond. Matter {\bf 19}, 065102 (2007).

\bibitem{MMS_2017} S. N. Majumdar, P. Mounaix, G. Schehr,
{\em Survival Probability of Random Walks and L\'evy Flights on a
Semi-Infinite Line}, J. Phys. A: Math. Theor. {\bf 50}, 465002 (2017).

\bibitem{Basu_18}
U. Basu, S. N. Majumdar, A. Rosso, G. Schehr, Phys. Rev. E {\bf 98}, 062121 (2018).

\bibitem{Sevilla_19}
F. J. Sevilla, A. V. Arzola, E. P. Cital, Phys. Rev. E {\bf 99}, 012145 (2019).

\bibitem{Dauchot_18}
O. Dauchot, V. D\'emery, preprint arXiv:1810.13303. 

\bibitem{Dhar_2019}
K. Malakar, A. Das, A. Kundu, K. V. Kumar, A. Dhar, preprint arXiv:1902.04171 . 

\bibitem{KMG}
S. Karlin, J. McGregor, Pacific J. Math. {\bf 9}, 1141 (1959).

\end{thebibliography}
\end{document}